\documentclass[11pt,a4paper]{scrartcl}
\usepackage[round]{natbib}

\usepackage[top=3cm, bottom=3cm, left=3cm, right=3cm]{geometry}

\usepackage{booktabs}

\usepackage{xcolor}
\usepackage{latexsym}              
\usepackage{amsmath}               
\usepackage{amssymb}               
\usepackage{amsfonts}              
\usepackage{amsthm}                
\usepackage{multirow}
\usepackage{tikz}                  
\usetikzlibrary{arrows,positioning,shapes}
\usepackage{tcolorbox}

\usepackage[english]{babel} 

\RequirePackage[%
  pdfstartview=FitH,%
  breaklinks=true,%
  bookmarks=true,%
  colorlinks=true,%
  linkcolor= blue,
  anchorcolor=blue,%
  citecolor=blue,
  filecolor=blue,%
  menucolor=blue,%
  urlcolor=blue%
  ]{hyperref}
  
  \AtBeginDocument{%
  \hypersetup{%
    pdfauthor={Florent Dewez, Benjamin Guedj, Arthur Talpaert and Vincent Vandewalle},%
  	urlcolor = blue,%
  	linkcolor = blue,%
  	citecolor = orange,%
    pdftitle={Title - compilation: \today}%
  }
}

\usepackage[mathcal]{eucal}
\usepackage{cleveref}
\crefname{assumption}{Assumption}{Assumptions}
\crefname{equation}{Eq.}{Eqs.}
\crefname{figure}{Fig.}{Figs.}
\crefname{table}{Table}{Tables}
\crefname{section}{Sec.}{Secs.}
\crefname{theorem}{Thm.}{Thms.}
\crefname{lemma}{Lemma}{Lemmas}
\crefname{corollary}{Cor.}{Cors.}
\crefname{example}{Example}{Examples}
\crefname{appendix}{Appendix}{Appendixes}
\crefname{remark}{Remark}{Remark}

\renewenvironment{proof}[1][\proofname]{{\bfseries #1.}}{\qed \\ }

\makeatother

\theoremstyle{plain}  
\newtheorem{theorem}{Theorem}[section]
\newtheorem{definition}[theorem]{Definition}

\newtheorem{lemma}[theorem]{Lemma}
\newtheorem{proposition}[theorem]{Proposition}

\newtheorem{remark}[theorem]{Remark}

\DeclareMathOperator*{\argmin}{arg\,min \:}
\DeclareMathOperator*{\rk}{\text{rank} \,}
\DeclareMathOperator*{\im}{\text{Im} \,}

\newcommand{\N}{\mathbb{N}}
\newcommand{\R}{\mathbb{R}}
\newcommand{\E}{\mathbb{E}}
\def\FF{\mathrm{FF}}
\def\TFC{\mathrm{TFC}}
\def\N{\mathrm{N1}}
\def\M{\mathrm{M}}
\def\MMO{\mathrm{MMO}}

\usepackage{arydshln}

\begin{document}

\title{An end-to-end data-driven optimisation framework for constrained trajectories}

\author{\textbf{Florent Dewez} \\ [2ex]
Inria, Lille - Nord Europe Research centre, France \\\\
\textbf{Benjamin Guedj} \\ [2ex]
Inria, Lille - Nord Europe Research centre, France \\
\emph{and} Centre for Artificial Intelligence,\\
Department of Computer Science,\\
University College London, United Kingdom \\\\
\textbf{Arthur Talpaert} \\ [2ex]
Inria, Lille - Nord Europe Research centre, France \\\\
\textbf{Vincent Vandewalle} \\ [2ex]
Inria, Lille - Nord Europe Research centre\\
\emph{and} Université de Lille, France\\
}
\date{}

\maketitle

\begin{abstract}
    Many real-world problems require to optimise trajectories under constraints. Classical approaches are based on optimal control methods but require an exact knowledge of the underlying dynamics, which could be challenging or even out of reach. In this paper, we leverage data-driven approaches to design a new end-to-end framework which is dynamics-free for optimised and realistic trajectories.
    We first decompose the trajectories on function basis, trading the initial infinite dimension problem on a multivariate functional space for a parameter optimisation problem. A maximum \emph{a posteriori} approach which incorporates information from data is used to obtain a new optimisation problem which is regularised. The penalised term focuses the search on a region centered on data and includes estimated linear constraints in the problem.
    We apply our data-driven approach to two settings in aeronautics and sailing routes optimisation, yielding commanding results. The developed approach has been implemented in the Python library PyRotor.\\
 
\end{abstract}

\tableofcontents


\section{Introduction} \label{sec:intro}
Many real-world problems require to optimise trajectories under constraints. The present paper stems from an initial work on aeronautics, and the quest for designing fuel efficient aircraft trajectories based on available flight data. We have reached a generic data-driven methodology which falls in the much broader field of trajectory optimisation under constraints. As such, it has potential applications to many real world problems, such as in robotics to minimise the work-based specific mechanical cost of transport \citep{sr2006} or in aerospace to reduce the total thermal flux when a space shuttle re-enters in the atmosphere \citep{trelat2012}.


In aeronautics, optimisation problems
are often formulated in terms of optimal control problems  \citep{dalmau2014, girardet2014, cots2018}. They can be solved by converting the problem into a 
parameter optimisation problem. This allows to take into account the dynamics of the system, leading to realistic solutions complying with additional constraints \citep[we refer to][for an overview]{Rao_asurvey}.

Nevertheless the differential equations describing the dynamics of the system of interest may be (partially) unknown. For instance, the differential system describing the motion of an aircraft moving in an air mass \citep{rommel2019} involves the lift and drag forces for which no analytic formulas exist. Aircraft manufacturer computes numerical models by means of heavy simulations and wind tunnel tests.
Another approach consists in reconstructing unknown forces based on physical formulas and available flight data; see for instance \citet{rommel2017, dewez2020industrywide} for results in aeronautics and \citet{RHCC2007} in the generic setting of parameter estimation for differential equations. However, while being promising, this reconstruction step requires restrictive assumptions and the statistical errors may impact strongly the solution of the optimal control problem. Moreover it does not tackle directly the optimisation problem.


The above approaches require intensive computations and may be affected by noise. The aim of our work is to provide realistic trajectories without involving complex and noisy dynamical systems. It is flexible enough and easily interpretable by experts while permitting the use of efficient optimisation algorithms. 
In the present work, we leverage available trajectory data to propose a thorough methodology which fulfils the above requirements. Our approach learns a model based on the observed trajectories, allowing in particular to extract some intrinsic properties in a data-driven way. It incorporates this modelling into an optimisation problem through a Bayesian approach. The resulting problem turns out to be constrained by the data in a simple and natural way. The main benefit on this approach is that it directly uses the information contained in the data, requiring no explicit information on the dynamics.

The methodology presented in this paper is specific to the situation where the user has access to trajectory data but, at the same time, the approach is intended to be generic enough so that it can be exploited in a wide range of applications. In particular it is certainly not restricted to the aeronautic setting.

Our approach first assumes that all the trajectories belong to a finite-dimensional space,
which allows to reduce the complexity of the problem with low information loss for a well-chosen basis.
In a Bayesian framework, we assume that the prior distribution of trajectories (through their related coefficients) is proportional to a decreasing exponential function of the cost, assuming that efficient trajectories are \emph{a priori} more likely than inefficient ones. The observed trajectories, that we call \emph{reference} trajectories, are interpreted as noisy observations of an efficient one, the noise following a centered Gaussian multivariate distribution. In a Bayesian perspective it is thus possible to deduce the \emph{posterior} distribution of the efficient trajectory given the reference trajectories. For the sake of simplicity, we focus on the mode of the \emph{posterior} distribution, the related objective function involves the cost of a trajectory and its squared Mahalanobis distance to a weighted average of reference trajectories. It can be interpreted as a penalised optimisation problem. 

The role of the penalisation in the objective function is to force the solution to be close to real trajectories. The strength of the penalisation is here controlled by a hyper-parameter and a tuning process is proposed to find an optimal balance between optimisation and (non-linear) constraints verification. Hence the optimised trajectory may inherit a realistic behaviour from the above closeness, even though the dynamics are not taken into account in our problem.
Further we note that this penalised term is actually quadratic and the optimisation problem is constrained by affine functions. So in certain cases, the problem is convex allowing to make use of very efficient algorithms.

A last remark on the penalised term is the fact that the underlying metrics is given by a covariance matrix estimated on the reference trajectories. It is noteworthy that this matrix not only indicates the most unconstrained directions for the optimisation but also reveals linear relations between variables. In particular, some of these relations may reflect the dynamics or may not be known by the user.

In a nutshell, this data-driven approach restricts the search space to a region centred on the data in a metric space reflecting features estimated from the data. In particular this property may help non-linear optimisation solvers to converge with a limited number of iterations.



\textbf{Outline. } We first describe our approach in \cref{sec:opti}. A brief description of the Python library we have develop is provided in \cref{sec:pyrotor}. \cref{sec:appli_aero} and \cref{sec:appli_nautic} are devoted to applications: the first one to the fuel reduction of aircraft during the climb and the second one to the maximisation of the work of a force field along a path. We finish the paper by discussing on future works to improve and generalise our optimisation methodology.

\section{An end-to-end optimisation workflow based on observed trajectories} \label{sec:opti}
We are interested in finding a trajectory $y^\star$ which minimises a certain cost function $F$, namely a solution of the following optimisation problem:
\begin{equation} \label{eq:opt_0}
    \widetilde{y}^\star \in \argmin_{y \in \mathcal{A}_\mathcal{G}(y_0, y_T)} F(y) \; .
\end{equation}
The set $\mathcal{A}_\mathcal{G}(y_0, y_T)$, which is defined in \cref{sec:2-1}, models the constraints the trajectory has to comply with, such that the initial and final conditions. Note that a trajectory is typically a multivariate function defined on an interval and its components are given by states and controls (which are not distinguished in this paper for the sake of presentation). If the constraints do not include the dynamics then the solution $\widetilde{y}^\star$ may be by far unrealistic. A strategy to force a more realistic pattern would be to add more user-defined constraints to the problem \eqref{eq:opt_0}, although this may be a complicated task and solving numerically the resulting problem could be computationally expensive.

In view of this, we provide in this section our full workflow to obtain a new optimisation problem which includes in a natural and simple way constraints coming from the data. This problem is actually designed to provide trajectories which have a realistic behaviour.

We begin with elementary but necessary definitions for trajectories and constraints in \cref{sec:2-1}. We aim at stating the optimisation problem in a finite basis space so we define in \cref{sec:2-2} the mathematical formalisation of how we decompose each trajectory as a projection on such a space. To extract information from the data for the optimisation problem, a statistical modelling on the projected space of the available trajectory data is done in \cref{subsec:ref_traj}. In \cref{subsec:modelling}, we put everything together to obtain our new optimisation problem \cref{subsec:modelling} via a maximum \emph{a posteriori} approach. \cref{subsec:quad_model} presents a handy computation regarding the cost function in a quadratic case, for the sake of completeness. Additional details can be found in the supplementary material.
\cref{subsec:iterations} focuses on a hyperparameter tuning for an optimal tradeoff between optimisation and additional (non-linear) constraints. Last but not least, \cref{sec:2-7} contains confidence intervals to assess the accuracy of the predicted optimised cost when the cost function is known up to a random noise term.

\subsection{Admissible trajectories modelling}
\label{sec:2-1}

We start with definitions.
\begin{definition}[Trajectory]
    Let $T > 0$ be a real number and let $D \geqslant 1$ be an integer. Any continuous $\R^D$-valued map $y$ defined on $[0,T]$, \emph{i.e.} $y \in \mathcal{C}\big([0,T], \R^D\big)$, is called a \emph{trajectory} over the time interval $[0,T]$. The $d$-th component of a trajectory $y$ will be denoted by $y^{(d)}$. As such, a trajectory is at least a continuous map on a finite interval.
\end{definition}
When optimising a trajectory with respect to a given criterion, the initial and final states are often constrained, that is to say the optimisation is performed in an affine subspace modelling these endpoints conditions. This subspace is introduced just below.

\begin{definition}[Endpoints conditions] \label{def:endpoints}
    Let $y_0, y_T \in \R^D$. We define the set $\mathcal{D}(y_0, y_T) \subset \mathcal{C}\big([0,T], \R^D\big)$ as follows:
    \begin{align*}
        y \in \mathcal{D}(y_0, y_T) \qquad
        & \Longleftrightarrow \qquad \left\{
		\begin{array}{l}
			y(0) = y_0 \\[2mm]
			y(T) = y_T
		\end{array} \right. \; .
    \end{align*}
\end{definition}

In many applications, the trajectories have to satisfy some additional constraints defined by a set of (nonlinear) functions. For instance these functions may model physical or user-defined constraints. In this paper, this set is not intended to include the dynamics of the system. We define now the set of trajectories verifying such additional constraints.

\begin{definition}[Additional constraints] \label{def:constraints}
    For $\ell = 1,\dots,L$, let $g_\ell$ be a real-valued function defined on $\R^D$. We define the set $\mathcal{G} \subset \mathcal{C}\big([0,T], \R^D\big)$ as the set of trajectories over $[0,T]$ satisfying the following $L$ inequality constraints given by the functions $g_\ell$, \emph{i.e.}
    \begin{equation*}
        y \in \mathcal{G} \qquad \Longleftrightarrow \qquad \forall \, \ell = 1, \dots, L \quad \forall \, t \in [0,T] \qquad g_\ell\big(y(t)\big) \leqslant 0 \; .
    \end{equation*}
\end{definition}

To finish we introduce the set of admissible trajectories which satisfy both the endpoints conditions and the additional constraints.

\begin{definition}[Admissible trajectory] \label{def:admissible}
    We define the set $\mathcal{A}_\mathcal{G}(y_0, y_T) \subset \mathcal{C}\big([0,T], \R^D\big)$ as follows:
    \begin{equation*}
        \mathcal{A}_\mathcal{G}(y_0, y_T) := \mathcal{D}(y_0, y_T) \cap \mathcal{G} \; .
    \end{equation*}
    Any element of $\mathcal{A}_\mathcal{G}(y_0, y_T)$ will be called an \emph{admissible trajectory}.
\end{definition}

\subsection{Projection for a finite-dimensional optimisation problem}
\label{sec:2-2}

In our approach, a theoretical optimisation problem in a finite-dimensional space is desired to reduce the inherent complexity of the problem. This can be achieved by decomposing the trajectories on a finite number of basis functions. While raw signals are unlikely to be described by a small number of parameters, this is not the case for smoothed versions of these signals which capture the important patterns. In particular, given a family of smoothed observed trajectories, one may suppose that there exists a basis such that the projection error on a certain number of basis functions of any trajectory is negligible.

From now on, the trajectories we consider are assumed to belong to a space spanned by a finite number of basis functions. For the sake of simplicity, we assume in addition that all the components of the trajectories can be decomposed on the same basis but with different dimensions. Extension to different bases is straightforward and does not change our findings but burdens the notation. 

\begin{definition} \label{def:projected_set}
    Let $\{\varphi_k\}_{k=1}^{+\infty}$ be an orthonormal basis of $L^2\big([0,T], \R \big)$ with respect to the inner product
    \begin{equation*}
        \langle f, g \rangle = \int_0^T f(t) \, g(t) \, dt \; ,
    \end{equation*}
    such that each $\varphi_k$ is continuous on $[0,T]$ and let $\mathcal{K} := \{K_d\}_{d=1}^D$ be a sequence of integers with $K := \sum_{d = 1}^D K_d$.
    We define the space of projected trajectories $\mathcal{Y}_\mathcal{K}(0,T) \subset \mathcal{C}\big([0,T], \R^D\big)$ over $[0,T]$ as
    \begin{equation*}
	    \mathcal{Y}_\mathcal{K}(0,T) := \prod_{d = 1}^D \text{span} \left\{ \varphi_{k} \right\}_{k = 1}^{K_d} \; .
    \end{equation*}
    If there is no risk of confusion, we write $\mathcal{Y}_\mathcal{K} := \mathcal{Y}_\mathcal{K}(0,T)$ for the sake of readability.
\end{definition}

\begin{remark}
    From the above definition, any projected trajectory $y \in \mathcal{Y}_\mathcal{K}$ is associated with a unique vector
    \begin{equation*}
        c = \Big( c_1^{(1)}, \dots, c_{K_1}^{(1)}, \; c_1^{(2)}, \dots, c_{K_2}^{(2)}, \dots, \; c_1^{(D)}, \dots, c_{K_D}^{(D)} \Big)^T \in \R^K
    \end{equation*}
    defined by
    \begin{equation} \label{eq:def_c}
        c_k^{(d)} := \big\langle y^{(d)}, \varphi_k \big\rangle = \int_0^T y^{(d)}(t) \, \varphi_{k}(t) \, dt \; .
    \end{equation}
    In other words, the vector $c$ is the image of the trajectory $y$ by the projection operator $\Phi : \mathcal{C}\big([0,T], \R^D \big) \longrightarrow \R^K$ defined by $\Phi y := c$, whose restriction $\Phi|_{\mathcal{Y}_\mathcal{K}}$ is bijective (as the Cartesian product of bijective operators). In particular, the spaces $\mathcal{Y}_\mathcal{K}$ and $\R^K$ are isomorphic, \emph{i.e.} $\mathcal{Y}_\mathcal{K} \simeq \R^K$.
\end{remark}

Regarding the endpoints conditions introduced in \cref{def:endpoints}, we prove in the following result that satisfying these conditions is equivalent to satisfying a linear system for a projected trajectory.

\begin{proposition} \label{prop:linear_const}
    A trajectory $y \in \mathcal{Y}_{\mathcal{K}}$ belongs to $\mathcal{D}(y_0, y_T)$ if and only if its associated vector $c := \Phi y \in \R^K$ satisfies the linear system
    \begin{equation} \label{eq:endpoint_syst}
        A(0,T) \, c = \Gamma \; ,
    \end{equation}
    where the matrix $A(0,T) \in \R^{2D \times K}$ and the vector $\Gamma \in \R^{2D}$ are defined as follows
    \begin{equation*}
        A(0, T) := \left(
        \begin{array}{ccccccc}
            \varphi_1(0) & \dots & \varphi_{K_1}(0) & & & & \\
            & & & \ddots & & & \\
            & & & & \varphi_1(0) & \dots & \varphi_{K_D}(0) \\
            \varphi_1(T) & \dots & \varphi_{K_1}(T) & & & & \\
            & & & \ddots & & & \\
            & & & & \varphi_1(T) & \dots & \varphi_{K_D}(T) \\
        \end{array} \right) \quad , \quad
        \Gamma := \left(
        \begin{array}{c}
            y_0\\
            y_T
        \end{array}
        \right) \; .
    \end{equation*}
\end{proposition}
\begin{proof}
    Let $y \in \mathcal{Y}_\mathcal{K}$ and let $c := \Phi y \in \R^K$. By the definition of the matrix $A(0,T)$, we have
    \begin{align*}
        A(0,T) \, c
            & = A(0,T) \Big( c_1^{(1)}, \dots, c_{K_1}^{(1)}, \; c_1^{(2)}, \dots, c_{K_2}^{(2)}, \dots, \; c_1^{(D)}, \dots, c_{K_D}^{(D)} \Big)^T \\
            & = \left( \sum_{k=1}^{K_1} c_k^{(1)} \varphi_k(0), \dots, \displaystyle \sum_{k=1}^{K_D} c_k^{(D)} \varphi_k(0), \dots, \sum_{k=1}^{K_1} c_k^{(1)} \varphi_k(T), \dots, \sum_{k=1}^{K_D} c_k^{(D)} \varphi_k(T) \right)^T \\
            & = \left(
        \begin{array}{c}
            y(0)\\
            y(T)
        \end{array}
        \right) \; .
    \end{align*}
    The conclusion follows directly from the preceding relation.
\end{proof}

\subsection{Reference trajectories modelling} \label{subsec:ref_traj}

Let us now suppose that we have access to $I$ recorded trajectories $y_{R_1},\ldots,y_{R_I}$, called \emph{reference trajectories}, coming from some experiments. We propose here a statistical modelling for these reference trajectories, permitting especially to exhibit some linear properties. This modelling will permit to take advantage of the information contained in these recorded trajectories when deriving optimisation problems in the next subsection.

These trajectories being recorded, they are in particular admissible and we assume that they belong to the space $\mathcal{Y}_{\mathcal{K}}(0,T)$. As explained previously they may be interpreted as smoothed versions of recorded signals. In particular each reference trajectory $y_{R_i}$ is associated with a unique vector $c_{R_i} \in \R^K$. Moreover we consider each reference trajectory as a noisy observation of a certain admissible and projected trajectory $y_*$. In other words we suppose that there exists a trajectory $y_* \in \mathcal{Y}_{\mathcal{K}} \cap \mathcal{A}_\mathcal{G}(y_0, y_T)$ associated with a vector $c_* \in \R^K$ satisfying
\begin{equation*}
    \forall \, i = 1, \dots, I \qquad c_{R_i} = c_* + \varepsilon_i \; .
\end{equation*}
The noise $\varepsilon_i$ is here assumed to be a centered Gaussian whose covariance matrix $\Sigma_i$ is of the form
\begin{equation*}
    \Sigma_i = \frac{1}{2 \omega_i} \, \Sigma \; ,
\end{equation*}
where $\Sigma \in \R^{K \times K}$. It is noteworthy that this matrix will not be known in most of the cases but an estimated covariance matrix can be computed on the basis of the reference vectors. The positive real numbers $\omega_i$ are here considered as weights so we require $\sum_{i=1}^I \omega_i = 1 \;$;
each $\omega_i$ plays actually the role of a noise intensity.
Further from the hypothesis that the trajectory $y_*$ and all the reference trajectories $y_{R_i}$ verify the same endpoints conditions, we deduce
\begin{equation*}
    A \, c_{R_i} = A \, c_* + A \, \varepsilon_i \qquad \Longleftrightarrow \qquad A \, \varepsilon_i = 0_{\R^{2D}} \qquad \Longleftrightarrow \qquad \varepsilon_i \in \ker A \; ,
\end{equation*}
for all $i = 1, \dots, I$ (we shorten $A(0,T)$ in $A$ when the context is clear). Hence the reference vector $c_*$ satisfies the following $I$ systems:
\begin{equation} \label{eq:model_c}
    \left\{ \begin{array}{l}
        c_{R_i} = c_* + \varepsilon_i \\[2mm]
        \varepsilon_i \sim \mathcal{N}(0_{\R^K}, \Sigma_i) \\[2mm]
        \varepsilon_i \in \ker A
    \end{array} \right. \; .
\end{equation}

To establish a more explicit system which is equivalent to the preceding one, we require the following preliminary proposition. Here we diagonalise the matrices $\Sigma$ and $A^T A$ by exploiting the fact that the image of the first one is contained in the null space of the other one and vice versa; this is shown in the proof. This property is actually a consequence of the above modelling: the endpoints conditions modelled by $A$ imply linear relations within the components of the vectors, which should be reflected by the covariance matrix $\Sigma$. The following result will be helpful to establish  \cref{prop:equiv_syst}.

\begin{proposition} \label{prop:matrix_structure}
    We define $\sigma := \rk \Sigma$ and $a := \rk A^T A$. In the setting of system \eqref{eq:model_c}, we have $\sigma + a \leqslant K$ and there exist an orthogonal matrix $V \in \R^{K \times K}$ and two matrices $\Lambda_\Sigma \in \R^{K \times K}$ and $\Lambda_A \in \R^{K \times K}$ of the following form:
    \begin{equation*}
        \Lambda_\Sigma = \left( \begin{array}{cc}
            \Lambda_{\Sigma,1} & 0_{\R^{\sigma \times (K-\sigma)}} \\[2mm]
            0_{\R^{(K-\sigma) \times \sigma}} & 0_{\R^{(K-\sigma) \times (K-\sigma)}}
        \end{array} \right) \qquad , \qquad 
        \Lambda_A = \left( \begin{array}{cc}
            0_{\R^{(K-a) \times (K-a)}} & 0_{\R^{(K-a) \times a}} \\[2mm]
            0_{\R^{a \times (K-a)}} & \Lambda_{A,2}
        \end{array} \right) \; ,
    \end{equation*}
    where $\Lambda_{\Sigma,1} \in \R^{\sigma \times \sigma}$ and $\Lambda_{A,2} \in \R^{a \times a}$ are diagonal matrices with positive elements, such that
    \begin{equation*}
        \Sigma = V \Lambda_\Sigma V^T \qquad , \qquad A^T A = V \Lambda_A V^T \; .
    \end{equation*}
\end{proposition}
\begin{proof}
    The starting point of the proof is to remark that we have
    \begin{equation} \label{eq:commute_zero}
        \Sigma \, A^T A = A^T A \, \Sigma = 0_{\R^{K \times K}} \; .
    \end{equation}
    Indeed using the hypothesis $\varepsilon_i \in \ker A$ for any $i=1,\dots,I$ gives
    \begin{equation*}
        \Sigma \, A^T A = 2 \omega_i \, \Sigma_i \, A^T A = 2 \omega_i \, \E(\varepsilon_i \varepsilon_i^T) \, A^T A = 2 \omega_i \, \E\big(\varepsilon_i \, (A \varepsilon_i)^T \big) \, A = 0_{\R^{K \times K}} \; ;
    \end{equation*}
    similar arguments prove the second equality in \eqref{eq:commute_zero}.
    First, we can deduce
    \begin{equation} \label{eq:inclusion}
        \im \Sigma \subseteq \ker A^T A \; ,
    \end{equation}
    which leads to $\sigma \leqslant K - a$ by the rank-nullity theorem. Equalities \eqref{eq:commute_zero} show also that $\Sigma$ and $A^T A$ are simultaneously diagonalisable (since they commute) so there exists an orthogonal matrix $V \in \R^{K \times K}$ such that
    \begin{equation} \label{eq:diag}
        \Sigma = V \Lambda_\Sigma V^T \qquad , \qquad A^T A = V \Lambda_A V^T \; ,
    \end{equation}
    where $\Lambda_\Sigma \in \R^{K \times K}$ and $\Lambda_A \in \R^{K \times K}$ are diagonal matrices. Permuting if necessary columns of $V$, we can write the matrix $\Lambda_\Sigma$ as follows:
    \begin{equation} \label{eq:lambda_sigma}
        \Lambda_\Sigma = \left( \begin{array}{cc}
            \Lambda_{\Sigma,1} & 0_{\R^{\sigma \times (K-\sigma)}} \\[2mm]
            0_{\R^{(K-\sigma) \times \sigma}} & 0_{\R^{(K-\sigma) \times (K-\sigma)}}
        \end{array} \right) \; ;
    \end{equation}
    in other words the $\sigma$ first column vectors of $V$ span the image of $\Sigma$. From the inclusion \eqref{eq:inclusion}, we deduce that these vectors belong to the null space of $A^T A$. Hence the $\sigma$ first diagonal elements of $\Lambda_A$ are equal to zero and, up to a permutation of the $K - \sigma$ last column vectors of $V$, we can write
    \begin{equation*}
        \Lambda_A = \left( \begin{array}{cc}
            0_{\R^{(K-a) \times (K-a)}} & 0_{\R^{(K-a) \times a}} \\[2mm]
            0_{\R^{a \times (K-a)}} & \Lambda_{A,2}
        \end{array} \right) \; ,
    \end{equation*}
    which ends the proof.
\end{proof}

\begin{remark}
    From equalities \eqref{eq:commute_zero}, we can also deduce
    \begin{equation*}
        \im A^T A \subseteq \ker \Sigma \; ,
    \end{equation*}
    showing that $\Sigma$ is singular. Consequently the Gaussian noise $\varepsilon_i$ involved in \eqref{eq:model_c} is degenerate.
\end{remark}

A new formulation of system \eqref{eq:model_c} which makes explicit the constrained and unconstrained parts of a vector satisfying this system is given in the following result. This is achieved by using the preceding result which allows to decompose the space $\R^K$ into three orthogonal subspaces. We prove that the restriction of the noise $\varepsilon_i$ to the first subspace is a non-degenerate Gaussian, showing that this first subspace corresponds to the unconstrained one. The two other subspaces describe affine relations coming from the endpoints conditions and from implicit relations within the vector components. These implicit relations, which may model for instance natural trends, are expected to be contained in the reference vectors $c_{R_i}$ and reflected by the (estimated) covariance matrix $\Sigma$.\\
Prior to this, let us write the matrix $V \in \R^{K \times K}$ introduced in \cref{prop:matrix_structure} as follows:
\begin{equation*}
    V = \big( V_1 \quad V_2 \quad V_3 \big) \; ,
\end{equation*}
where $V_1 \in \R^{K \times \sigma}$,  $V_2 \in \R^{K \times K-\sigma-a}$ and  $V_3 \in \R^{K \times a}$. We emphasise that the column-vectors of the matrices $V_1$ and $V_3$ do not overlap according to the property $\sigma + a \leqslant K$ proved in \cref{prop:matrix_structure}. In particular the matrix $V_2$ has to be considered only in the case $\sigma + a < K$. Further for any $c \in \R^K$, we will use the notations 
\begin{equation*}
    \widetilde{c} := V^T c \qquad, \qquad \widetilde{c}_\ell := V_\ell^T c \; ,
\end{equation*}
for $\ell=1,2,3$. Finally we consider the singular value decomposition of $A$ coming from the diagonalisation of the symmetric matrix $A^T A$ with $V$:
\begin{equation*}
    A = U S_A V^T \; ,
\end{equation*}
where $U \in \R^{2D \times 2D}$ is orthogonal and $S_A \in \R^{2D \times K}$ is a rectangular diagonal matrix of the following form:
\begin{equation} \label{eq:s_a}
    S_A = \big( 0_{\R^{2D \times K-2D}} \quad S_{A,2} \big) \; ,
\end{equation}
with $S_{A,2} := \sqrt{\Lambda_{A,2}} \in \R^{2D \times 2D}$.
    
\begin{proposition} \label{prop:equiv_syst}
    Suppose that the matrix $A$ is full rank, \emph{i.e.} $a = 2D$. Then for any $i=1, \dots, I$, system \eqref{eq:model_c} is equivalent to the following one:
    \begin{equation} \label{eq:model_c_tilde_2}
        \left\{ \begin{array}{l}
            \widetilde{c}_{R_i, 1} = \widetilde{c}_{*,1} + \widetilde{\varepsilon}_{i, 1} \\[2mm]
            \displaystyle \widetilde{\varepsilon}_{i, 1} \sim \mathcal{N} \left(0_{\R^\sigma}, \frac{1}{2 \omega_i} \, \Lambda_{\Sigma,1} \right) \\[3mm]
            \widetilde{c}_{*,2} = V_2^T c_{R_i} \\[2mm]
            \widetilde{c}_{*,3} = \, S_{A,2}^{-1} \, U^T \Gamma
        \end{array} \right. \; .
    \end{equation}
\end{proposition}
\begin{proof}
    We first prove that system \eqref{eq:model_c} is equivalent to
    \begin{equation} \label{eq:model_c_tilde}
        \left\{ \begin{array}{l}
            \widetilde{c}_{R_i} = \widetilde{c}_* + \widetilde{\varepsilon}_i \\[2mm]
            \displaystyle \widetilde{\varepsilon}_i \sim \mathcal{N} \left(0_{\R^K}, \frac{1}{2 \omega_i} \, \Lambda_\Sigma \right) \\[3mm]
            S_A \, \widetilde{c}_* = U^T \Gamma
        \end{array} \right. \; .
    \end{equation}
    The matrix $V$ being orthogonal, it is non-singular and so we have for all $i = 1, \dots, I$,
    \begin{equation*}
        c_{R_i} = c_* + \varepsilon_i \qquad \Longleftrightarrow \qquad \widetilde{c}_{R_i} = \widetilde{c}_* + \widetilde{\varepsilon}_i \; ,
    \end{equation*}
    and, since $\Sigma_i = \frac{1}{2 \omega_i} \, \Sigma = \frac{1}{2 \omega_i} \, V \Lambda_\Sigma V^T$, we obtain
    \begin{equation*}
        \varepsilon_i \sim \mathcal{N}(0_{\R^K}, \Sigma_i) \qquad \Longleftrightarrow \qquad \widetilde{\varepsilon}_i \sim \mathcal{N} \left(0_{\R^K}, \frac{1}{2 \omega_i} \, \Lambda_\Sigma \right) \; .
    \end{equation*}
    Finally the property $\varepsilon_i \in \ker A$ is equivalent to
    \begin{align*}
        A \, c_* = \Gamma \qquad 
            & \Longleftrightarrow \qquad U S_A V^T c_* = \Gamma \\
            & \Longleftrightarrow \qquad S_A \, \widetilde{c}_* = U^T \Gamma \; ,
    \end{align*}
    proving that the systems \eqref{eq:model_c} and \eqref{eq:model_c_tilde} are equivalent.
    Now the fact that the $K-\sigma$ last diagonal elements of $\Lambda_\Sigma$ are zero implies that the components $\widetilde{c}_{*,2} \in \R^{K-\sigma-2D}$ and $\widetilde{c}_{*,3} \in \R^{2D}$ are constant. From the first equality of \eqref{eq:model_c_tilde}, we have on one side
    \begin{equation*}
        \widetilde{c}_{R_i, 2} = \widetilde{c}_{*,2} \qquad \Longleftrightarrow \qquad V_2^T c_{R_i} = \widetilde{c}_{*,2} \; ,
    \end{equation*}
    for any $i = 1, \dots, I$. On the other side, combining the last relation of the system \eqref{eq:model_c_tilde} with the form of the matrix $S_A$ given in \eqref{eq:s_a} permits to obtain
    \begin{align*}
        S_A \, \widetilde{c}_* = U^T \Gamma \qquad
            & \Longleftrightarrow \qquad S_{A,2} \, \widetilde{c}_{*,3} = U^T \Gamma \nonumber \\
            & \Longleftrightarrow \qquad \widetilde{c}_{*,3} = \, S_{A,2}^{-1} \, U^T \Gamma \; ,
    \end{align*}
    the last equivalence being justified by the hypothesis that the matrix $A$ is full rank (which implies that the diagonal matrix $S_{A,2}$ is non-singular).
\end{proof}

The above decomposition gives us access to non-degenerated density of $\widetilde{c}_{R_i, 1}$ given  $\widetilde{c}_{*,1}$ which is later denoted by $u(\widetilde{c}_{R_i, 1}|\widetilde{c}_{*,1})$. In next section we will assume a prior distribution on $\widetilde{c}_{*,1}$ with high density for low values of the cost function $F$.

\subsection{A trajectory optimisation problem via a Maximum A Posteriori approach} \label{subsec:modelling}

Before introducing the Bayesian framework, let first recall that we are interested in minimising a certain cost function $F: \mathcal{C}\big([0,T], \R^D \big) \longrightarrow \R$ over the set of projected and admissible trajectories $\mathcal{Y}_\mathcal{K} \cap \mathcal{A}_\mathcal{G}(y_0, y_T)$. As explained previously, we propose here a methodology leading to a constrained optimisation problem based on the reference trajectories and designed to provide realistic trajectories.
Technically speaking, we seek for the mode of a \emph{posterior} distribution which contains information from the reference trajectories. The aim of this subsection is then to obtain the \emph{posterior} distribution via Bayes's rule, using in particular the precise modelling of the reference trajectories given in \cref{prop:equiv_syst} and defining an accurate prior distribution with high density for low values of the cost function $F$.

To do so, we recall firstly that all the trajectories considered here are assumed to belong to the space $\mathcal{Y}_\mathcal{K}$ which is isomorphic to $\R^K$. So each trajectory is here described by its associated vector in $\R^K$, permitting in particular to define distributions over finite-dimensional spaces. We recall also that the reference trajectories are interpreted as noisy observations of a certain $y_*$ associated with a $c_*$. According to \cref{prop:equiv_syst}, this vector complies with some affine conditions which are described by the following subspace $\mathcal{V}_1$:
\begin{equation} \label{eq:v1}
    c \in \mathcal{V}_1 \qquad \Longleftrightarrow \qquad \left\{
    \begin{array}{l}
        V_2^T c = V_2^T c_{R_i} \\[2mm]
        V_3^T c = S_{A,2}^{-1} \, U^T \Gamma
    \end{array}
    \right. \; .
\end{equation}
Hence a vector $c$ belonging to $\mathcal{V}_1$ is described only through its component $\widetilde{c}_1 := V_1^T c$. In addition we note that the definition of $\mathcal{V}_1$ does not depend actually on the choice of $i$ since $V_2^T c_{R_i}$ has been proved to be constant in \cref{prop:equiv_syst}. Further we emphasise that the matrix $A$ is supposed to be full rank in this case and we have $\mathcal{V}_1 \simeq \R^\sigma$; we recall that $\sigma$ is the rank of the covariance matrix $\Sigma$.

Let us now define the cost function $F$ over the spaces $\R^K$ and $\mathcal{V}_1$. This is necessary to define the \emph{prior} distribution and to establish our optimisation problem.

\begin{definition}[Cost functions] \label{def:cost}
    Let $\check{F}: \R^K \longrightarrow \R$ and $\widetilde{F}: \R^\sigma \longrightarrow \R$ be the functions defined by
    \begin{itemize}
        \item $\displaystyle \check{F}(c) := F \big( \Phi|_{\mathcal{Y}_\mathcal{K}}^{-1} c \big)$ ;
        \item $\displaystyle \widetilde{F}(\widetilde{c}_1) := F\Big( \Phi|_{\mathcal{Y}_\mathcal{K}}^{-1} \, V \Big( \widetilde{c}_1^T \quad c_{R_i}^{\ T} V_2 \quad \Gamma^T U \, \big(S_{A,2}^{-1} \big)^T \Big)^T \Big)$ .
    \end{itemize}
\end{definition}

\begin{remark}
    From the preceding definition, we observe that for any $y \in \mathcal{Y}_\mathcal{K}$ and its associated vector $c \in \R^K$, we have
    \begin{equation*}
        \check{F}(c) = F \big( \Phi|_{\mathcal{Y}_\mathcal{K}}^{-1} c \big) = F(y) \; .
    \end{equation*}
    Further for any $c \in \mathcal{V}_1$, we have
    \begin{align*}
        \check{F}(c) = F \big( \Phi|_{\mathcal{Y}_\mathcal{K}}^{-1} c \big) = F \big( \Phi|_{\mathcal{Y}_\mathcal{K}}^{-1} V \widetilde{c} \big) = F\Big( \Phi|_{\mathcal{Y}_\mathcal{K}}^{-1} \, V \Big( \widetilde{c}_1^T \quad c_{R_i}^{\ T} V_2 \quad \Gamma^T U \, \big(S_{A,2}^{-1} \big)^T \Big)^T \Big) = \widetilde{F}(\widetilde{c}_1) \; .
    \end{align*}
    We deduce that $\widetilde{F}$ is actually the restriction of $\check{F}$ to the subspace $\mathcal{V}_1$.
\end{remark}

From now on, the trajectory $y_*$ and the associated vector $c_*$ will be considered as random variables and will be denoted by $y$ and $c$. We are interested in the \emph{posterior} distribution 
\begin{equation*}
    u(\widetilde{c}_1 \, | \, \widetilde{c}_{R_1, 1}, \dots, \widetilde{c}_{R_I, 1}) \; ,
\end{equation*}
which depends only on the free component $\widetilde{c}_1$ of $c \in \mathcal{V}_1$, the two other ones $\widetilde{c}_2$ and $\widetilde{c}_3$ being fixed according to \eqref{eq:v1}. We use Bayes's rule to model the \emph{posterior} via the \emph{prior} and likelihood distributions, leading to
\begin{equation*}
    u(\widetilde{c}_1 \, | \, \widetilde{c}_{R_1, 1}, \dots, \widetilde{c}_{R_I, 1}) \propto u(\widetilde{c}_{R_1, 1}, \dots, \widetilde{c}_{R_I, 1} \, | \, \widetilde{c}_1) \, u(\widetilde{c}_1) \; .
\end{equation*}
Assuming now that the vectors $\widetilde{c}_{R_i, 1}$ are independent gives
\begin{equation*}
    u(\widetilde{c}_{R_1, 1}, \dots, \widetilde{c}_{R_I, 1} \, | \, \widetilde{c}_1) \, u(\widetilde{c}_1) = \prod_{i=1}^I u(\widetilde{c}_{R_i, 1} \, | \, \widetilde{c}_1) \, u(\widetilde{c}_1) \; .
\end{equation*}
The above likelihood is given by the modelling of the reference trajectories detailed in \cref{prop:equiv_syst}. In this case, we have
\begin{equation*}
    u(\widetilde{c}_{R_i, 1} \, | \, \widetilde{c}_1) \propto \exp\Big( -\omega_i \, \big( \widetilde{c}_1 - \widetilde{c}_{R_i, 1} \big)^T \Lambda_{\Sigma,1}^{-1} \big( \widetilde{c}_1 - \widetilde{c}_{R_i, 1} \big) \Big) \; .
\end{equation*}
The prior distribution is obtained by assuming that the most efficient trajectories (with respect to the cost function) are \emph{a priori} the most likely ones:
\begin{equation} \label{eq:model_prior}
    u(\widetilde{c}_1) \propto \exp \Big( -\kappa^{-1} \widetilde{F}(\widetilde{c}_1) \Big) \; ,
\end{equation}
where $\kappa > 0$.
Putting everything together and taking the negative of the logarithm gives the following minimisation problem, whose solution is the Maximum \emph{A Posteriori} estimator:
\begin{equation} \label{eq:opt_proj}
    \left\{
    \begin{array}{l}
         \displaystyle \widetilde{c}_1^\star \in \argmin_{\widetilde{c}_1 \in \R^\sigma} \widetilde{F}(\widetilde{c}_1) + \kappa \sum_{i=1}^I \omega_i \, \big( \widetilde{c}_1 - \widetilde{c}_{R_i, 1} \big)^T \Lambda_{\Sigma,1}^{-1} \big( \widetilde{c}_1 - \widetilde{c}_{R_i, 1} \big) \\[2mm]
         \widetilde{c}_2 = V_2^T c_{R_i} \\[2mm]
         \widetilde{c}_3 = S_{A,2}^{-1} \, U^T \Gamma
    \end{array} \; ,
    \right.
\end{equation}
where $i$ is arbitrarily chosen in $\{1, \dots, I \}$.

Let us now rewrite the above optimisation problem with respect to the variable $c = V \widetilde{c} \in \R^K$ in order to make it more interpretable.
\begin{proposition} \label{prop:opt_equiv}
    The optimisation problem \eqref{eq:opt_proj} is equivalent to the following one:
    \begin{equation} \label{eq:opt_proj_2}
        c^\star \in \argmin_{c \in \mathcal{V}_1} \check{F}(c) + \kappa \sum_{i=1}^I \omega_i \, \big( c - c_{R_i} \big)^T \Sigma^\dagger \big( c - c_{R_i} \big) \; ,
    \end{equation}
    where $\Sigma^\dagger \in \R^{K \times K}$ denotes the pseudoinverse of the matrix $\Sigma$.
\end{proposition}
\begin{proof}
    From \eqref{eq:lambda_sigma}, we deduce
    \begin{align*}
        \sum_{i=1}^I \omega_i \, \big( \widetilde{c}_1 - \widetilde{c}_{R_i, 1} \big)^T \Lambda_{\Sigma,1}^{-1} \big( \widetilde{c}_1 - \widetilde{c}_{R_i, 1} \big)
            & = \sum_{i=1}^I \omega_i \, \big( \widetilde{c} - \widetilde{c}_{R_i} \big)^T \Lambda_{\Sigma}^\dagger \big( \widetilde{c} - \widetilde{c}_{R_i} \big) \\
            & = \sum_{i=1}^I \omega_i \, \big( c - c_{R_i} \big)^T \, V \Lambda_{\Sigma}^\dagger V^T \, \big( c - c_{R_i} \big) \\
            & = \sum_{i=1}^I \omega_i \, \big( c - c_{R_i} \big)^T \Sigma^\dagger \big( c - c_{R_i} \big) \; .
    \end{align*}
    And from the proof of \cref{prop:equiv_syst}, we have 
    \begin{equation*}
        A \, c = \Gamma \qquad \Longleftrightarrow \qquad \widetilde{c}_3 = \, S_{A,2}^{-1} \, U^T \Gamma \; ,
    \end{equation*}
    proving that $c \in \mathcal{V}_1$.
\end{proof}

To conclude, let us comment on this optimisation problem.
\begin{enumerate}
    \item To interpret the optimisation problem \eqref{eq:opt_proj_2} (or equivalently \eqref{eq:opt_proj}) from a geometric point of view, let us consider the following new problem:
    \begin{equation} \label{eq:primal}
        \begin{array}{ll}
            & \displaystyle \min_{\widetilde{c}_1 \in \R^\sigma} \widetilde{F}(\widetilde{c}_1) \\[2mm]
            \text{s.t. } & \displaystyle \sum_{i=1}^I \omega_i \, \big( \widetilde{c}_1 - \widetilde{c}_{R_i, 1} \big)^T \Lambda_{\Sigma,1}^{-1} \big( \widetilde{c}_1 - \widetilde{c}_{R_i, 1} \big) \leqslant \widetilde{\kappa}
        \end{array}
    \end{equation}
    where $\lambda \geqslant 0$. Here we suppose that $\widetilde{F}$ is strictly convex and that the problem \eqref{eq:primal} has a solution (which is then unique). By Slater's theorem \citep[Subsec. 5.2.3]{convexopt}, the strong duality holds for the problem \eqref{eq:primal}. It can then be proved that there exists a certain $\lambda^\star \geqslant 0$ such that the solution of \eqref{eq:primal} is the minimiser of the strictly convex function
    \begin{equation*}
        \widetilde{c}_1 \longmapsto \widetilde{F}(\widetilde{c}_1) + \lambda^\star \sum_{i=1}^I \omega_i \, \big( \widetilde{c}_1 - \widetilde{c}_{R_i, 1} \big)^T \Lambda_{\Sigma,1}^{-1} \big( \widetilde{c}_1 - \widetilde{c}_{R_i, 1} \big) \; ,
    \end{equation*}
    which is actually the objective function of the optimisation problem \eqref{eq:opt_proj} for $\kappa = \lambda^\star$. Hence the problem \eqref{eq:opt_proj} minimises the cost $\widetilde{F}$ in a ball centered on the weighted average of the reference trajectories. In particular if the reference trajectories are close to an optimal one with respect to $\widetilde{F}$ then one could expect the solution of \eqref{eq:opt_proj} to be equal to this optimal trajectory.
    \item Further the optimisation problem \eqref{eq:opt_proj_2} takes into account the endpoints conditions through the subspace $\mathcal{V}_1$ but not the additional constraints. However as explained in the preceding point, the solution is close to realistic trajectories and so is likely to comply with the additional constraints for a well-chosen parameter $\kappa > 0$. We refer to \cref{subsec:iterations} for more details on an iterative method for the tuning of $\kappa$. In particular, a right choice for this parameter is expected to provide an optimised trajectory with a realistic behaviour. This is for instance illustrated in \cref{sec:appli_aero}.
    \item Taking into account the linear information from the available data through the covariance matrix $\Sigma$ allows to restrict the search to the subspace $\mathcal{V}_1$ describing these relations. This is of particular interest when implicit relations (modelled by the sub-matrix $V_2$) are revealed by the estimation of $\Sigma$ on the basis of the reference trajectories; in this case, these implicit relations may not be known by the expert.
    \item The optimisation problem \eqref{eq:opt_proj_2} has linear constraints and a quadratic penalised term. For instance, if the cost function $\check{F}$ is a convex function then we obtain a convex problem for which efficient algorithms exist.
\end{enumerate}

\subsection{Quadratic cost for a convex optimisation problem} \label{subsec:quad_model}

In this short subsection, we focus on a particular case where the cost function $F$ is defined as the integral of an instantaneous quadratic cost function, \emph{i.e.}  
\begin{equation} \label{eq:quad_f}
    \forall \, y \in \mathcal{C}\big([0,T], \R^D \big) \qquad F(y) = \int_0^T f(y(t)) \, dt \; ,
\end{equation}
where $f : \R^D \longrightarrow \R$ is quadratic.
Even though such a setting may appear to be restrictive, we emphasise that quadratic models may lead to highly accurate approximations of variables, as it is illustrated in \cref{sec:appli_aero}. For a quadratic instantaneous cost, the associated function $\check{F}: \R^K \longrightarrow \R$ can be proved to be quadratic as well and can be explicitly computed. In the following result, we provide a quadratic optimisation problem equivalent to \eqref{eq:opt_proj_2}.

\begin{proposition} \label{prop:quad_min_pb}
    Suppose that the cost function $F$ is of the form \eqref{eq:quad_f} with $f$ quadratic. Then the optimisation problem \eqref{eq:opt_proj_2} is equivalent to the following one:
    \begin{equation} \label{eq:quad_min_pb}
        c^\star \in \argmin_{c \in \mathcal{V}_1} c^T \Big( \check{Q} + \kappa \, \Sigma^\dagger \Big) c + \left( \check{w} - 2 \kappa \sum_{i=1}^I \omega_i \, \Sigma^\dagger c_{R_i} \right)^{\hspace{-5pt}T} c \; ,
    \end{equation}
    where $\check{Q} \in \R^{K \times K}$ and $\check{w} \in \R^K$ can be explicitly computed from $f$.
\end{proposition}
\begin{proof}
    We defer the proof to the supplementary material.
\end{proof}

In particular this permits to derive sufficient conditions on the parameter $\kappa > 0$ so that the optimisation problem is proved to be equivalent to a quadratic program \citep[Sec. 4.4]{convexopt}, namely the objective function is convex quadratic together with affine constraints. In practice, this allows to make use of efficient optimisation libraries to solve numerically \eqref{eq:quad_min_pb}.

\subsection{Iterative process to comply with additional constraints} \label{subsec:iterations}

As explained in \cref{subsec:modelling}, the trajectory optimisation problem \eqref{eq:opt_proj_2} is constrained by the endpoints conditions and by implicit linear relations revealed by the reference trajectories. Nevertheless the additional constraints introduced in \cref{def:constraints} are not taken into account in this problem. In practice such constraints assure that natural or user-defined features are verified and so a trajectory which does not comply with these constraints may be considered as unrealistic.

Our aim is then to assure that the trajectory $y^\star = \Phi|_{\mathcal{Y}_\mathcal{K}}^{-1} c^\star$, where $c^\star \in \mathcal{V}_1$ is the solution of the optimisation problem \eqref{eq:opt_proj_2}, verifies the additional constraints, \emph{i.e.} belongs to the set $\mathcal{G}$. A first solution would be to add the constraint $\Phi|_{\mathcal{Y}_\mathcal{K}}^{-1} c \in \mathcal{G}$ in the optimisation problem \eqref{eq:opt_proj_2}. However depending on the nature of the constraints functions $g_\ell$, this may lead to nonlinear constraints which could be costly from a numerical point of view. The solution we propose consists rather in exploiting the degree of freedom coming from the parameter $\kappa > 0$ appearing in the problem \eqref{eq:opt_proj_2}.

First of all, let us factorise the problem \eqref{eq:opt_proj_2} by $\kappa$ to obtain the following new one for the sake of presentation:
\begin{equation} \label{eq:opt_proj_3}
    c^\star \in \argmin_{c \in \mathcal{V}_1} \nu \, \check{F}(c) + \sum_{i=1}^I \omega_i \, \big( c - c_{R_i} \big)^T \Sigma^\dagger \big( c - c_{R_i} \big) \; ,
\end{equation}
where $\nu := \kappa^{-1}$. On one hand, we observe that the solution of the optimisation problem \eqref{eq:opt_proj_3} for the limit case $\nu = 0$ is given by $\sum_{i=1}^I \omega_i \, c_{R_i}$ which is the average of the reference vectors. In this case, one may expect that the associated average trajectory complies with the constraints but is unlikely to optimise the cost function $F$. On the other hand, for very large $\nu > 0$, the second term of the objective function in \eqref{eq:opt_proj_3} can be considered as negligible compared to the first one. In this case, the cost of the solution will surely be smaller than the costs of the reference trajectories but no guarantee regarding the additional constraints can be established in a general setting.

Given these observations, the task is then to find an appropriate value $\nu^\star > 0$ in order to reach a trade-off between optimising and remaining close to the reference trajectories to comply with the additional constraints. Many methods can be developed to find such a $\nu^\star$ and, among those based on iterative processes, linear or binary search algorithms can be considered. In this case, one has to set firstly a maximal value $\nu_{max}$ so that the solution of \eqref{eq:opt_proj_3} with $\nu_{max}$ is unlikely to satisfy the constraints and to perform secondly the search over the interval $(0,\nu_{max})$. Since the solution for $\nu = 0$ is assumed to be admissible, we expect that the binary search will find a $\nu^\star > 0$ leading to an optimised trajectory belonging to $\mathcal{G}$.

\subsection{Confidence bounds on the integrated cost}
\label{sec:2-7}

In practice the cost function $F$ considered is an estimation of the true cost $F^{\star}$, a random variable which cannot be fully predicted based on $y$. If the distribution $F(y)$ would be known it would be possible to deduce confidence bound on $F^{\star}$. This is for instance possible by considering multivariate functional regression \citep{RHCC2007}. 

The simplest case from the estimation point of view is to consider that $F^{\star}$ is the integral of some instantaneous consumption function $f^{\star}$
as in \cref{subsec:quad_model}, and to estimate the parameters of the standard multivariate regression 
$$
f^{\star}(y(t)) = f(y(t)) + \varepsilon(t), 
$$
where the random noise $\varepsilon(t)$ is assumed to follow a centered Gaussian distribution with variance $\sigma$. In this case $F^{\star}$ can be expressed as the integral of a stochastic process
$$
F^{\star}(y) := \int_{0}^{T} f^{\star}(y(t)) \, dt \;  = F(y) + \int_{0}^{T} \varepsilon(t) \, dt \;.  
$$
then assuming that $(\varepsilon(t))_{t \in [0,T]}$ independent we obtain
$$
\int_{0}^{T} \varepsilon(t) \, dt \; \sim \mathcal{N}(0,T \sigma^2).
$$
Thus $F^{\star}(y)$ follows a Gaussian distribution centered on $F(y)$ and with variance equals to $T\sigma^2$. This makes it possible to compute confidence bounds on $F^{\star}(y)$. For a confidence level $1-u$, $u\in [0,1]$, a confidence interval for $F^{\star}(y)$ is obtained as
$$ \texttt{CI}^{1-u}(F^{\star}(y)) = F(y) \pm \zeta_{1-\frac{u}{2}} 
\sqrt{T}\sigma,$$
where $\zeta_{1-\frac{u}{2}}$ is the quantile of order $1-\frac{u}{2}$ of the standard Gaussian distribution.




The assumption that $f$ and $\sigma^2$ are known is relevant since they  are estimated based on a huge amount of training data. The assumption of white Gaussian noise can be seen as unrealistic, however it appears to be the only route to explicit calculus. A more complex strategy could be derived using Gaussian processes, which is beyond the scope of this paper.






\section{The Python library Pyrotor} \label{sec:pyrotor}
The above optimisation methodology is aimed at being used in a wide range of applications, from path planning for industrial robots \citep{chettibi2004703} to fuel-efficient aircraft trajectories \citep{dewez2020industrywide, rommel2019}. We therefore contribute a generic Python library \texttt{PyRotor} (standing for \textbf{Py}thon \textbf{Ro}ute \textbf{t}rajectory \textbf{o}ptimise\textbf{r}) which is intended to a large audience. In particular, this library has been used to obtain the numerical results given in the two following sections.

When using the \texttt{PyRotor} library, the practitioner has to define the endpoints conditions as a dictionary, the additional constraints in a list of functions, the name of the basis and the dimension for each variable. The current version of the library covers only the case of \cref{subsec:quad_model}, that is to say the cost is given by a quadratic instantaneous function. This permits to make use of \cref{prop:quad_min_pb} in which a quadratic objective function is given. We mention that future releases of \texttt{PyRotor} are intended to cover more general cost functions. The value of the parameter $\nu_{max}$ in \eqref{eq:opt_proj_3} can also be manually set depending on the application. The Legendre basis is currently the only basis implemented in the first version of \texttt{PyRotor} (via the \texttt{legendre} module from \texttt{NumPy} package \citep{2020NumPy-Array}) but future developments including other general bases are planned. Further the user indicates a path to a directory containing the data, each reference trajectory being contained in a \texttt{csv} file. The covariance matrix $\Sigma$ is here estimated by using the \texttt{sklearn.covariance} package from the \texttt{Python} library \texttt{scikit-learn} \citep{scikit-learn}. Two optimisation solvers are proposed: the generic solver \texttt{minimize(method=’trust-constr’)} \citep{Trust-region} from \texttt{SciPy} software \citep{2020SciPy-NMeth} and the quadratic programming solver from \texttt{CVXOPT} software \citep{cvxopt}. The latter is intended to speed up the execution in case of convex quadratic objective function. Once the arguments are given by the user, a class is created and the optimisation is performed by executing a method from this class. At the end, the optimised trajectory is provided in a dataframe: at each time, the position of the trajectory is given together with the value of $f$. The total cost is also computed and a quantitative comparison in terms of savings with the reference trajectories can be also displayed.

The open source \texttt{PyRotor} library is developed on GitHub and welcomes contributions from its users: we favour a community-based development to foster the diffusion of our work towards practitioners.
\texttt{PyRotor} is intended to be PEP8 compliant and purposely rely on high standard coding practices. The continuous development platform \href{https://travis-ci.com}{Travis} is used to certify the latest builds of the library. Finally we provide \href{https://jupyter-notebook.readthedocs.io/en/stable/}{Jupyter notebooks} for examples on how to use \texttt{PyRotor} along with online documentation. \texttt{PyRotor} is available at
\href{https://github.com/bguedj/pyrotor}{https://github.com/bguedj/pyrotor}.


\section{Application 1: trajectory optimisation for fuel-efficient aircrafts} \label{sec:appli_aero}
In this section, we consider the aeronautic problem of reducing the total fuel consumption of an aircraft during the climb phase. This example illustrates the key role played by the reference trajectories since we are able to obtain realistic and optimised trajectories thanks to a simple modelling involving few constraints.

\subsection{Modelling} \label{subsec:appli_aero_model}

Here the trajectories are supposed to be in a vertical plane and are defined by the altitude $h$, the Mach number $\M$ and the engines rotational speed $\N$ (expressed as a percentage of a maximal value). Hence a trajectory $y$ in this setting is a continuous $\R^3$-valued map defined on $[0,T]$, where $T$ is a maximal climb duration fixed by the user. Hence we have
\begin{equation*}
    \forall \, t \in [0,T] \qquad y(t) := \big( h(t), \M(t), \N(t) \big) \; .
\end{equation*}
The quantity to minimise is the total fuel consumption $\TFC: \mathcal{C}\big( [0,T], \R^3 \big) \longrightarrow \R_+$ which is defined via the fuel flow $\FF: \R^3 \longrightarrow \R_+$ as follows\footnote{In the notation of \cref{subsec:quad_model}, $\FF$ and $\TFC$ play respectively the role of $f$ and $F$.}:
\begin{equation*}
    \TFC(y) := \int_0^T \FF\big(y(t)\big) \, dt \; .
\end{equation*}
Regarding the endpoints conditions, we require the trajectory to start at the altitude $h_0$ with Mach number $\M_0$ and to end at the altitude $h_T$ with Mach number $\M_T$. In particular, the reference trajectories we use have to verify these conditions.

We consider also additional constraints which are conventional in the aeronautic setting:
\begin{itemize}
    \item The rate of climb, \emph{i.e.} the time-derivative of the altitude, has to be upper bounded by a given maximal value $\gamma_{max}$ during the whole climb;
    \item The Mach number should not exceed a certain value called the maximum operational Mach ($\MMO$).
\end{itemize}
The final time of the climb is given by $T^\star \in [0,T]$ which is the first time where the aircraft reaches $h_T$ with Mach number $\M_T$.

Finally we mention that the fuel flow model $\FF$ is here estimated. To do so, we exploit the reference trajectories which contain recorded altitude, Mach number, engines power and fuel flow for each second of the flight. Having access to these data, we are in position to fit a statistical model. Following the numerical results in \cite{dewez2020industrywide} which show that polynomials can accurately model aeronautic variables, we consider a polynomial model of degree 2 for the fuel flow. In particular the requirements for the cost function in the current version of \texttt{PyRotor} are fulfilled. The prediction accuracy of the resulting estimated model is assessed in the following subsection.

\subsection{Numerical results}

We present now numerical results based on real flight data for the above aeronautic problem. Here we have access to 2,162 recorded short and medium-haul flights performed by the same narrow-body airliner type, provided by a partner airline. In particular they can not be publicly released for commercial reasons. The data is here recorded by the Quick Access Recorder (QAR). 

Before considering the optimisation setting, we estimate a fuel flow model specific to the climb phase and to the considered airliner type. To do so we extract the signals of the four variables of interest (altitude, Mach number, engines rotational speed and fuel flow) and keep the observations from the take-off to the beginning of the cruise without level-off phases. Smoothing splines are then applied to the raw signals to remove the noise. We sample each 5 seconds to reduce the data set size without impacting strongly the accuracy of the resulting models. At the end, we obtain 494,039 observations which are randomly split into training and test sets to fit a polynomial model of degree 2 using the \texttt{scikit-learn} library. The RMSE and MAPE values of this model on the test set are respectively equal to $3.64 \times 10^{-2}$ kg.s$^{-1}$ and 1.73\%.

Regarding the optimisation, we are interested in climb phases from 3,000~ft to 38,000~ft. We mention that we remove lower altitudes because operational procedures constraint heavily the trajectory during the very beginning of the climb. Further the initial and final Mach numbers are required to be equal to 0.3 and 0.78. It is noteworthy that the optimisation solvers used in \texttt{PyRotor} allow linear inequality conditions, permitting to slightly relax the endpoints conditions. Here we tolerate an error of 100~ft for the altitude and an error of 0.01 for the Mach number. The initial and final $\N$ values are let unconstrained. Finally the $\MMO$ and $\gamma_{max}$ are respectively set to 0.82 and 3,600~ft.min$^{-1}$.

The reference trajectories are given by 48 recorded flights which satisfy the above climb endpoints conditions among the 2,162 available ones. All these selected flights are used to estimate the covariance matrix involved in the optimisation problem. On the other hand, we use only the 5 most fuel-efficient flights in the objective function to focus on a domain containing the most efficient recorded flights. Further the maximal duration $T$ is here fixed to the duration of the longest climb among the 5 most fuel-efficient ones we use.

Legendre polynomials are used as the functional basis spanning the space in which lies the trajectories. Since we consider narrow-body airliners, polynomials are expected to be relevant to describe the slow variations of such aircrafts. Here the dimensions associated with the altitude, the Mach number and the engines power are given respectively by 4, 10 and 6. The reference vectors $c_{R_i}$ are then computed using the formula \eqref{eq:def_c}. At the end, we amount to solving a constrained optimisation problem in a space of dimension 20. 

We are then in position to apply the optimisation method developed in \cref{sec:opti} using the \texttt{PyRotor} library. First of all a relevant value for $\nu_{max} > 0$ has to be fixed. In order to propose a realistic optimised climb, we choose a $\nu_{max}$ relatively small so that the optimised climb remains close to the reference ones. In particular, the quadratic objective function in \eqref{eq:opt_proj_3} turns out to be convex for all $\nu \in [0,\nu_{max}]$ permitting to use the quadratic programming solver from \texttt{CVXOPT} software imported in \texttt{PyRotor}. The preprocessing of the reference trajectories and the optimisation steps have been executed 100 times using \texttt{PyRotor} on an Intel Core i7 6 cores running at 2.2~GHz. The mean of the execution time for both steps is equal to 3.76~s with standard deviation 0.11~s, illustrating that the library is time-efficient in this setting.

\begin{figure}
    \centering
    \includegraphics[width=\textwidth]{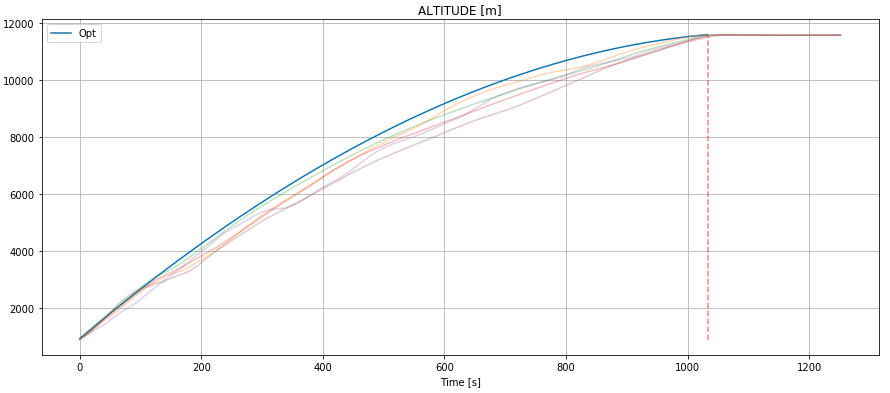}
    \includegraphics[width=\textwidth]{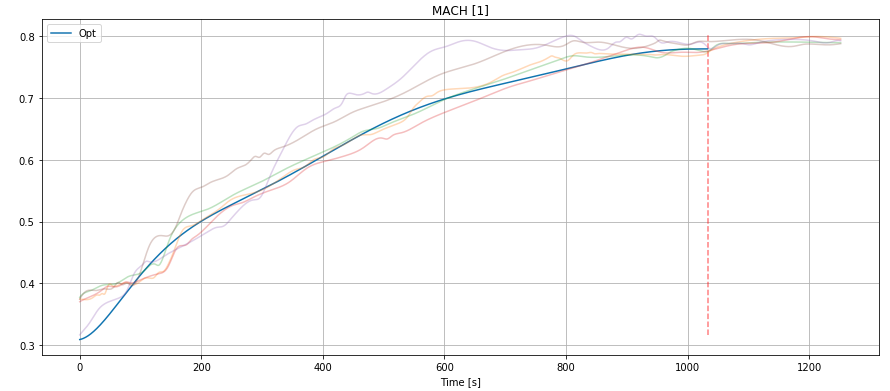}
    \includegraphics[width=\textwidth]{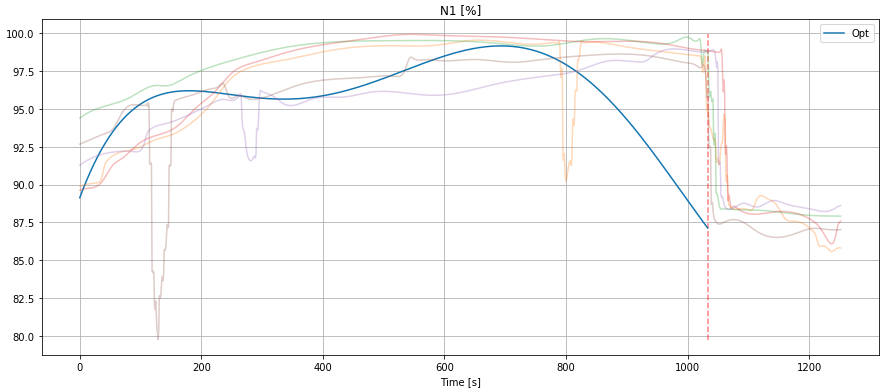}
    \caption{Optimised and reference altitudes, Mach numbers and engines rotational speeds -- The optimised trajectory is represented by the blue curves.}
    \label{fig:opti_climb}
\end{figure}

A plot of the optimised trajectory obtained using \texttt{PyRotor} is given in \cref{fig:opti_climb}. We observe that the optimised trajectory seeks to reach the maximum altitude in the minimum amount of time; this is in accordance with the existing literature (see for instance \citet{dalmau2014} and references therein). In particular, the duration $T^\star$ is equal to 1,033 seconds which is actually slightly shorter than the reference durations. We note also that the optimised Mach number shares a very similar pattern with the references. On the other hand, the optimised engines rotational speed tends to slowly decrease until the cruise regime before reaching the top of climb. This is not the case for the reference engines speed which falls to the cruise regime just after reaching the final altitude. Most of the savings seem to be achieved in these last moments of the climb. At last but not least, the optimised trajectory presents a realistic pattern inherited from the reference trajectories.

For a quantitative comparison, we refer to \cref{tab:savings} which provides statistical information on the fuel savings. The mean savings 16.54\% together with the fact that the optimised trajectory verifies the additional constraints show that these first results are promising, motivating further studies. For instance one could model environmental conditions or take into account Air Traffic Control constraints for more realistic modellings.

\begin{table}
    \caption{\label{tab:savings}Statistical description of the fuel savings of the optimised trajectory -- The savings are compared with the 48 recorded flights satisfying the present endpoints and the total consumption of the optimised trajectory is estimated by using the statistical model for the fuel flow - $Q_1$, $Q_2$ and $Q_3$ refer to the first, second and third quartiles.}
    \centering
    \fbox{%
    \begin{tabular*}{\textwidth}{@{\extracolsep{\fill}} l*{6}{c}r}
        & Mean & Std & Min & $Q_1$ & $Q_2$ & $Q_3$ & Max \\
        \hline
        Fuel savings [kg] & 260.38 & 86.21 & 71.79 & 202.40 & 261.87 & 330.32 & 393.73 \\
        Percentage [\%] & 16.54 & 4.73 & 5.27 & 13.56 & 16.88 & 20.39 & 23.39
    \end{tabular*}}
\end{table}


\section{Application 2: trajectory optimisation to maximise work of a force field} \label{sec:appli_nautic}
Here we consider the following generic example: given a moving point in a force field, find a trajectory starting and ending at two different given points which maximises the work of the force along the trajectory while minimising the travelled distance. For instance, this corresponds to a very simple modelling of a sailing boat which seeks to increase the power of the wind at each time, \emph{i.e.} maximising the wind work, without travelling a too large distance. This second example demonstrates that our generic optimisation approach is flexible enough to take into account derivatives of trajectories and hence to cover dynamics settings.

\subsection{Modelling} \label{subsec:appli_phys_nautic}

To model this problem, we suppose without loss of generality that the trajectories are defined on the (time-)interval $[0,1]$ and we let  $V: \R^D \longrightarrow \R^D$ denote a vector field. Furthermore the trajectories are assumed here to be continuously differentiable, \emph{i.e.} they belong to $\mathcal{C}^1\big([0,1], \R^D \big)$. The work of $V$ along a trajectory $y \in \mathcal{C}^1\big( [0,1], \R^D \big)$ is defined as
\begin{equation*}
    W(y, \dot{y}) := \int_0^1 V\big(y(t)\big)^T \dot{y}(t) \, dt \; ;
\end{equation*}
here $\dot{y}$ denotes the derivative of $y$ with respect to the independent variable $t$. Moreover using Hamilton's principle in Lagrangian mechanics, it can be shown that the trajectory with constant velocity (\emph{i.e.} a straight line travelled at constant speed) is the minimum of the following functional,
\begin{equation*}
    J(\dot{y}) = \int_0^1 \big\| \dot{y}(t) \big\|_2^2 \, dt \; ,
\end{equation*}
where the starting and ending points of $y$ are fixed and different. This functional can be then used to control the travelled distance. It follows that minimising the cost function
\begin{equation*}
    F_\alpha(y, \dot{y}) := \alpha J(\dot{y}) - W(y, \dot{y}) = \int_0^1 \alpha \big\| \dot{y}(t) \big\|_2^2 - V\big(y(t)\big)^T \dot{y}(t) \, dt \; ,
\end{equation*}
where $\alpha \geqslant 0$ is arbitrarily chosen, is expected to lead to an optimised trajectory reflecting a trade-off between maximising the work and minimising the distance. Further we require the trajectory to stay in the hypercube $[0,1]^D$ and to start and to end respectively at $y_0 \in [0,1]^D$ and $y_1 \in [0,1]^D$.

Now we remark that the above cost function involves the (time-)derivative $\dot{y}$. So one has to derive a formula permitting to compute the derivative of any trajectory $y = \Phi|_{\mathcal{Y}_\mathcal{K}}^{-1} c \in \mathcal{Y}_\mathcal{K}$ from its associated vector $c \in \R^K$, especially to compute $\check{F}(c)$. For instance, this can be easily achieved by assuming that each element of the functional basis is continuously differentiable. Indeed we can differentiate in this case any $y \in \mathcal{Y}_\mathcal{K}$:
\begin{equation*}
    \forall \, d = 1, \dots, D \qquad \dot{y}^{(d)} = \sum_{k=1}^{K_d} c_k^{(d)} \dot{\varphi}_k =  \left( \frac{d}{dt} \Phi|_{\mathcal{Y}_\mathcal{K}}^{-1} c \right)^{(d)} \; .
\end{equation*}
We deduce then the following formula for $\check{F}(c)$ in the present setting:
\begin{equation*}
    \check{F}(c) := F_\alpha\left( \Phi|_{\mathcal{Y}_\mathcal{K}}^{-1} c, \frac{d}{dt} \Phi|_{\mathcal{Y}_\mathcal{K}}^{-1} c  \right) \; .
\end{equation*}
Here the vector $c$ contains information on both position and velocity, permitting especially to keep the problem dimension unchanged. To finish, let us remark that it is possible to make the above formula for $\check{F}$ explicit with respect to $c$ in certain settings. For instance it is possible to derive an explicit quadratic formula for $\check{F}(c)$ when the integrand defining $F_\alpha$ is quadratic with respect to $y(t)$ and $\dot{y}(t)$; this formula is implemented in \texttt{PyRotor} and the arguments to obtain it are similar to those proving \cref{prop:quad_min_pb}.

\subsection{Numerical results}

Numerical results based on randomly generated data for the above physical application are presented in this section.

First of all we consider trajectories with two components $y^{(1)}$ and $y^{(2)}$ lying in the square $[0,1]^2$ for the sake of simplicity. We set the starting and ending points as follows:
\begin{equation*}
    y^{(1)}(0) = 0.111 \quad , \quad y^{(2)}(0) = 0.926 \quad , \quad y^{(1)}(1) = 0.912 \quad , \quad y^{(2)}(1) = 0.211 
\end{equation*}
with a tolerated error $1 \times 10^{-4}$, and the vector field $V: \R^2 \longrightarrow \R^2$ is here defined by
\begin{equation*}
    V\Big(x^{(1)}, x^{(2)}\Big) = \Big( 0, x^{(1)} \Big)^T \; .
\end{equation*}
Given the above endpoints and the vector field, we observe that the force modelled by $V$ will be in average a resistance force to the motion. Indeed the force is oriented toward the top of the square while the moving point has to go downward. Further let us note that the integrand of the cost function $F_\alpha$ in the present setting is actually quadratic with respect to $y(t)$ and $\dot{y}(t)$, so that an explicit quadratic formula for $\check{F}(c)$ implemented in \texttt{PyRotor} is available.

Here the reference trajectories are obtained through a random generation process. To do so, we define an arbitrarily trajectory $y_R$ verifying the endpoints conditions and we compute its associated vector $c_R$; Legendre polynomials are once again used and the dimensions of $y^{(1)}$ and $y^{(2)}$ are here set to 4 and 6. Let us note that $y_R$ is designed in such a way that it has a relevant pattern but not the optimal one. Then we construct a set of reference trajectories by adding centered Gaussian noises to $c_R$. It is noteworthy that the noise is generated in such a way that it belongs to the null space of the matrix $A$ describing the endpoints conditions; the resulting noised trajectories satisfy then these conditions. Further the trajectories which go out of the square $[0,1]^2$ are not kept. At the end, we get 122 generated reference trajectories assumed to be realistic in this setting, each of them containing 81 time observations. Among these reference trajectories, we use the 10 most efficient ones with respect to the cost $F_\alpha$.

In the present example, we set a $\nu_{max}$ relatively large to explore a large domain around the reference trajectories. In this case, the objective function of the optimisation problem \eqref{eq:opt_proj_3} may be not convex even if it is still quadratic. So we make use of the generic optimisation solver \texttt{minimize(method=’trust-constr’)} imported in \texttt{PyRotor}. Regarding the execution time, we have randomly and uniformly generated 100 values in the interval $[0,10]$ for the parameter $\alpha$ and executed \texttt{PyRotor} for each of them. The mean of \texttt{PyRotor} execution time is 0.44~s with standard deviation 0.03~s on an Intel Core i7 6 cores running at 2.2~GHz.

\begin{figure}[t]
    \centering
    \includegraphics[width=\textwidth]{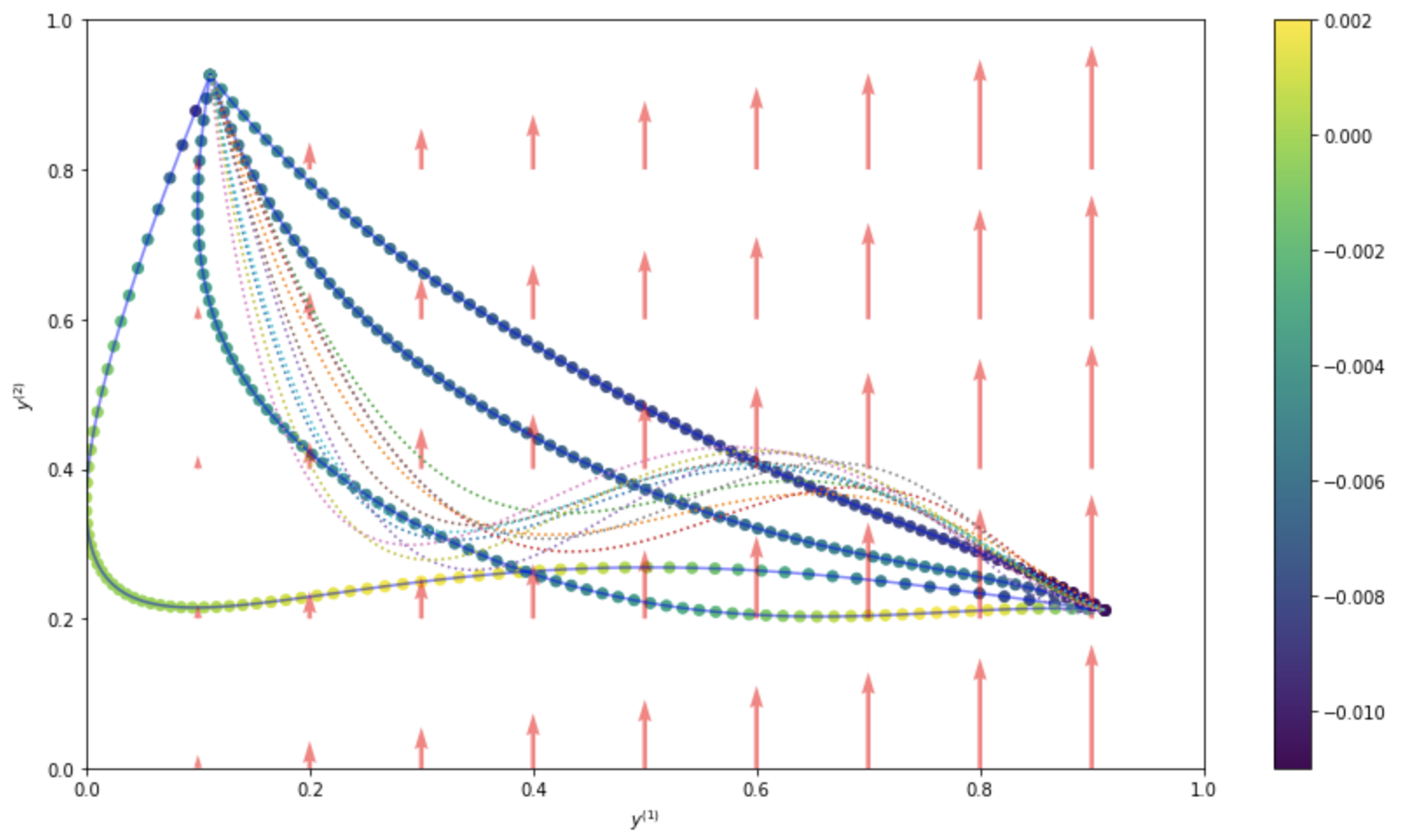}
    \caption{Optimised trajectories in the square $[0,1]^2$ for $\alpha \in \{0, 0.35, 1, 10 \}$ -- Optimised and reference trajectories are respectively given by plain and dotted curves -- Coloured dots indicate the power value of the force at different points of the optimised trajectories and the bar shows the scale -- Red arrows represent the pattern of the vector field $V$.}
    \label{fig:opti_work_velocity}
\end{figure}

In \cref{fig:opti_work_velocity}, we plot 4 optimised trajectories associated with different values of $\alpha$: 0, 0.35, 1 and 10. As expected the trajectory associated with the largest value of $\alpha$ gives the most straight trajectory while the most curvy one is associated with $\alpha = 0$. In particular, the latter tends to move to the left at the beginning where the force $V$ is the smallest before going to the ending point in a quasi-straightforward way so that the force is perpendicular to the motion. This example illustrates especially that our optimisation approach may lead to optimised trajectories which differ from the reference ones to reduce more the cost.

A quantitative comparison in terms of work gains for different values of $\alpha$ is provided in \cref{tab:max_work}. The results confirm the above observations on the curves and show that a right value for $\alpha$ has to be fixed depending on the setting.

\begin{table}
    \caption{\label{tab:max_work}Statistical description of the work gains in percentage for $\alpha \in \{0, 0.35, 1, 10 \}$ -- The values have been computed by using the 122 available reference trajectories -- Negative percentages indicate that no work gains have been obtained -- $Q_1$, $Q_2$ and $Q_3$ refer to the first, second and third quartiles.}
    \centering
    \fbox{
    \begin{tabular*}{\textwidth}{@{\extracolsep{\fill}} l*{6}{c}r}
        & Mean & Std & Min & $Q_1$ & $Q_2$ & $Q_3$ & Max \\
        \hline
        $\alpha = 0$ & 73.43 & 2.36 & 68.63 & 71.90 & 73.25 & 74.67 & 80.69 \\
        $\alpha = 0.35$ & 45.88 & 4.81 & 36.09 & 42.75 & 45.49 & 48.39 & 60.66 \\
        $\alpha = 1$ & $-6.12$ & 9.43 & $-25.31$ & $-12.26$ & $-6.88$ & $-1.20$ & 22.87 \\
        $\alpha = 10$ & $-34.54$ & 11.96 & $-58.87$ & $-42.32$ & $-35.50$ & $-28.30$ & 2.22 \\
    \end{tabular*}}
\end{table}


\section{Conclusion / Discussion}
We have proposed an approach for data-driven trajectories  optimisation without involving dynamical system. The approach can work based on a known cost function or a cost function learnt from the data. The modelling of the trajectories allows to take into account explicit and implicit linear constraints on the coefficients  in the optimisation problem. Contrary to full optimisation approaches, our method finds a trade-off between high density  constraints-compliant solutions and fully optimised solutions through the tuning of the regularisation. In the aeronautic framework our approach leads to promising fuel-efficient trajectories. Our approach is generic enough to be applied to other physical settings such as the motion of a moving point in a force field (such as a sailing boat).

Some perspectives of this work are first to further exploit the flexibility of Bayesian setting, by not only searching for the mode of the \emph{posterior} distribution but also sampling by means of MCMC algorithms. A second perspective would be to consider a clustering of reference trajectories, and apply our strategy on each cluster, then particularise the optimal trajectory depending on the cluster. Last but not least,  we aim at adapting our approach where some component of the trajectory would be categorical variables: this would be particularly useful for decision making processes in various disciplines.

\section*{Acknowledgements}
This project has received funding from the Clean Sky 2 Joint Undertaking under the European Union’s Horizon 2020 research and innovation programme under grant agreement No 815914 (PERF-AI). The funder had no role in study design, data collection and analysis, decision to publish, or preparation of the manuscript. Data that support the findings of this study have been gathered from the Quick Access Recorder (QAR) of aircrafts operated by a private airline, hence can not be released publicly for commercial reasons. The authors are grateful to Baptiste Gregorutti and Pierre Jouniaux for fruitful discussions about the modelling of the problem and the validation of the results for the aeronautic application.

\appendix
\section{Appendix}


\def\FF{\mathrm{FF}}
\def\TFC{\mathrm{TFC}}
\def\N{\mathrm{N1}}
\def\M{\mathrm{M}}
\def\MMO{\mathrm{MMO}}

Here we focus on the case where the cost function $F: \mathcal{C}\big([0,T], \R^D\big) \longrightarrow \R^D$ is of the following form
\begin{equation} \label{eq:quad_F}
    F(y) = \int_0^T y(t)^T Q \, y(t) + w^T y(t) + r \, dt \; ,
\end{equation}
where $Q \in \R^{D \times D}$ is symmetric, $w \in \R^D$ and $r \in \R$. In this setting, we provide explicit formulas for the costs $\check{F}: \R^K \longrightarrow \R$ and $\widetilde{F}: \R^\sigma \longrightarrow \R$ defined in Section 2.4. A sufficient condition on the parameter $\kappa > 0$ so that the optimisation problem
\begin{equation} \label{eq:opt_ref}
    c^\star \in \argmin_{c \in \mathcal{V}_1} \check{F}(c) + \kappa \sum_{i=1}^I \omega_i \, \big( c - c_{R_i} \big)^T \Sigma^\dagger \big( c - c_{R_i} \big) \; ,
\end{equation}
is a quadratic program in the present setting is then derived. From Section 2.4, we recall that the preceding optimisation problem is equivalent to
\begin{equation} \label{eq:opt_ref_tilde}
    \left\{
    \begin{array}{l}
         \displaystyle \widetilde{c}_1^\star \in \argmin_{\widetilde{c}_1 \in \R^\sigma} \widetilde{F}(\widetilde{c}_1) + \kappa \sum_{i=1}^I \omega_i \, \big( \widetilde{c}_1 - \widetilde{c}_{R_i, 1} \big)^T \Lambda_{\Sigma,1}^{-1} \big( \widetilde{c}_1 - \widetilde{c}_{R_i, 1} \big)  \\[2mm]
         \widetilde{c}_2 = V_2^T c_{R_i} \\[2mm]
         \widetilde{c}_3 = S_{A,2}^{-1} \, U^T \Gamma
    \end{array} \; .
    \right.
\end{equation}

\begin{lemma} \label{lem:ff}
    Suppose that the cost function $F$ is of the form \eqref{eq:quad_F}. Then the costs $\check{F}$ and $\widetilde{F}$ are quadratic and explicit formulas are given in \eqref{eq:check_F} and \eqref{eq:widetilde_F}.
\end{lemma}
\begin{proof}
    Let $c \in \R^K$ and let $y := \Phi|_{\mathcal{Y}_\mathcal{K}}^{-1} c \in \mathcal{Y}_\mathcal{K}$ be its associated trajectory, which can be represented as follows:
    \begin{equation*}
        \forall \, d \in \{1, \dots, D\} \qquad y^{(d)} = \sum_{k=1}^{K_d} c_k^{(d)} \, \varphi_k \; .
    \end{equation*}
    We also remark that each component of the vector
    \begin{equation*}
        c = \Big( c_1^{(1)}, \dots, c_{K_1}^{(1)}, \; c_1^{(2)}, \dots, c_{K_2}^{(2)}, \dots, \; c_1^{(D)}, \dots, c_{K_D}^{(D)} \Big)^T
    \end{equation*}
    can be simply described by a single parameter so that we can write $c = (c_1, c_2, \dots, c_K)^T$.
    \begin{itemize}
        \item \underline{Computation of $\check{F}$:}\\
        We first insert the preceding representation of $y$ into the above quadratic integrand to obtain:
        \begin{align}
    	    & y(t)^T Q \, y(t) + w^T y(t) + r \nonumber \\
    		& \hspace{1cm} = \sum_{d_1, d_2 = 1}^D \sum_{k_1 = 0}^{K_{d_1}} \sum_{k_2 = 0}^{K_{d_2}} Q_{d_1 d_2} \, c_{k_1}^{(d_1)} c_{k_2}^{(d_2)} \, \varphi_{k_1}(t) \varphi_{k_2}(t) + \sum_{d = 1}^D \sum_{k=0}^{K_d} w_d \, c_k^{(d)} \, \varphi_k(t) + r \; , \label{eq:double_sum}
        \end{align}
        for all $t \in [0,T]$. The next step of the proof consists in changing the indices of the above sums. To do so, let us define the matrix $\overline{Q} \in \R^{K \times K}$ and the vector $\overline{w} \in \R^K$ as
        \begin{equation*}
            \overline{Q} := \left(
            \begin{array}{ccc}
                Q_{11} \, J_{K_1, K_1} & \ldots & Q_{1D} \, J_{K_1, K_D} \\
                \vdots & & \vdots \\
                Q_{D1} \, J_{K_D, K_1} & \ldots & Q_{DD} \, J_{K_D, K_D}
            \end{array} \right) \; , \quad
            \overline{w} := \big( w_1 \, J_{1, K_1} \quad \dots \quad w_D \, J_{1, K_D} \big)^T \; ,
        \end{equation*}
        where $J_{m,n}$ is the all-ones matrix of size $m \times n$. We also introduce the map $\overline{\varphi} \in \mathcal{C}\big([0,T], \R^K\big)$ as
        \begin{equation*}
            \overline{\varphi}(t) := \Big( \varphi_1(t), \dots, \varphi_{K_1}(t), \; \varphi_1(t), \; \dots, \varphi_{K_2}(t), \dots, \; \varphi_1(t), \dots, \varphi_{K_D}(t) \Big)^T \; ,
        \end{equation*}
        for all $t \in [0,T]$, where the $\varphi_k$ are the functional basis elements. We are now in position to change the indices in the sums appearing in \eqref{eq:double_sum}:
        \begin{align*}
            & \sum_{d_1, d_2 = 1}^D \sum_{k_1 = 0}^{K_{d_1}} \sum_{k_2 = 0}^{K_{d_2}} Q_{d_1 d_2} \, c_{k_1}^{(d_1)} c_{k_2}^{(d_2)} \, \varphi_{k_1}(t) \varphi_{k_2}(t) + \sum_{d = 1}^D \sum_{k=0}^{K_d} w_d \, c_k^{(d)} \, \varphi_k(t) + r \\
            & \hspace{1cm} = \sum_{k_1, k_2 = 1}^{K} \overline{Q}_{k_1 k_2} \, c_{k_1} c_{k_2} \, \overline{\varphi}_{k_1}(t) \overline{\varphi}_{k_2}(t) + \sum_{k = 1}^{K} \overline{w}_k \, c_k \, \overline{\varphi}_k(t) + r \; ,
        \end{align*}
        where we have used the above rewriting of the vector $c$. Integrating finally over $[0,T]$ gives
        \begin{align}
            \check{F}(c)
                & = \int_0^T y(t)^T Q \, y(t) + w^T y(t) + r \, dt \nonumber \\
    		    & = \sum_{k_1, k_2 = 1}^{K} \overline{Q}_{k_1 k_2} \int_0^T \overline{\varphi}_{k_1}(t) \, \overline{\varphi}_{k_2}(t) \, dt \, c_{k_1} c_{k_2} + \sum_{k = 1}^{K} \overline{w}_k \int_0^T \overline{\varphi}_k(t) \, dt \, c_k + r T \nonumber \\
    		    & = \sum_{k_1, k_2 = 1}^{K} \check{Q}_{k_1 k_2} \, c_{k_1} c_{k_2} + \sum_{k = 1}^{K} \check{w}_k \, c_k + r T \nonumber\\
    		    & = c^T \check{Q} c + \check{w}^T c + r T \; , \label{eq:check_F}
        \end{align}
        where we have defined
        \begin{equation} \label{eq:widetilde_q_w}
	        \check{Q}_{k_1 k_2} :=  \overline{Q}_{k_1 k_2} \int_0^T \overline{\varphi}_{k_1}(t) \, \overline{\varphi}_{k_2}(t) \, dt \qquad , \qquad \check{w}_k := \overline{w}_k \int_0^T \overline{\varphi}_k(t) \, dt \; .
        \end{equation}
        \item \underline{Computation of $\widetilde{F}$:}\\
        By the definition of $\widetilde{F}$ given in Section 2.4, we have
        \begin{equation*}
            \widetilde{F}(\widetilde{c}_1) = \check{F}\Big( V \big( \widetilde{c}_1^T \quad \widetilde{c}_{2,3}^{\ T} \big)^T \Big) \; ,
        \end{equation*}
        where $V$ has been introduced in Section 2.3 and $\widetilde{c}_{2,3} \in \R^{K-\sigma}$ is defined as follows:
        \begin{equation*}
            \widetilde{c}_{2,3} := \left(
            \begin{array}{c}
                V_2^T c_{R_i} \\[2mm]
                S_{A,2}^{-1} U^T \Gamma
            \end{array} \right) \; ,
        \end{equation*}
        here the index $i$ is arbitrarily chosen in $\{1, \dots, I \}$ since the vector $V_2^T c_{R_i}$ has been proved to be independent from $i$. We introduce now the matrix $\check{Q}_V := V^T \check{Q} V$ and the vector $\check{w}_V := V^T \check{w}$ which can be decomposed as follows:
        \begin{equation*}
            \check{Q}_V = \left( \begin{array}{cc}
                \check{Q}_{V, 11} & \check{Q}_{V, 12} \\[2mm]
                \check{Q}_{V, 21} & \check{Q}_{V, 22}
            \end{array} \right) \quad , \quad
            \check{w}_V = \left( \begin{array}{c}
                \check{w}_{V, 1} \\[2mm]
                \check{w}_{V, 2} 
            \end{array} \right) \; ,
        \end{equation*}
        where $\check{Q}_{V, 11} \in \R^{\sigma \times \sigma}$, $\check{Q}_{V, 12} \in \R^{\sigma \times (K - \sigma)}$, $\check{Q}_{V, 21} \in \R^{(K-\sigma) \times \sigma}$, $\check{Q}_{V, 22} \in \R^{(K-\sigma) \times (K-\sigma)}$, $\check{w}_{V, 1} \in \R^\sigma$ and $\check{w}_{V, 2} \in \R^{K-\sigma}$. Given this and the preceding point, we obtain
        \begin{align*}
            \widetilde{F}(\widetilde{c}_1)
                & = \big( \widetilde{c}_1^T \quad \widetilde{c}_{2,3}^{\ T} \big) \big( V^T \check{Q} V \big) \big( \widetilde{c}_1^T \quad \widetilde{c}_{2,3}^{\ T} \big)^T + \big( V^T \check{w} \big)^T \big( \widetilde{c}_1^T \quad \widetilde{c}_{2,3}^{\ T} \big)^T + r T \\
                & = \big( \widetilde{c}_1^T \quad \widetilde{c}_{2,3}^{\ T} \big) \check{Q}_V \big( \widetilde{c}_1^T \quad \widetilde{c}_{2,3}^{\ T} \big)^T + \check{w}_V^T \big( \widetilde{c}_1^T \quad \widetilde{c}_{2,3}^{\ T} \big)^T + r T \\
                & = \widetilde{c}_1^T \, \check{Q}_{V, 11} \, \widetilde{c}_1 + \widetilde{c}_1^T \, \check{Q}_{V, 12} \, \widetilde{c}_{2,3} + \widetilde{c}_{2,3}^{\ T} \, \check{Q}_{V, 21} \, \widetilde{c}_1 + \widetilde{c}_{2,3}^{\ T} \, \check{Q}_{V, 22} \, \widetilde{c}_{2,3} \\
                & \hspace{1cm} + \check{w}_{V, 1}^{\ T} \, \widetilde{c}_{1} + \check{w}_{V, 2}^{\ T} \, \widetilde{c}_{2,3} + r T \; .
        \end{align*}
        Rearranging the preceding terms and using the fact that $\check{Q}_V$ is symmetric gives
        \begin{equation} \label{eq:widetilde_F}
            \widetilde{F}(\widetilde{c}_1) = \widetilde{c}_1^T \, \widetilde{Q} \, \widetilde{c}_1 + \widetilde{w}^T \widetilde{c}_1 + \widetilde{r} \; ,
        \end{equation}
        where
        \begin{align}
            & \bullet \quad \widetilde{Q} := \check{Q}_{V, 11} \; ; \label{eq:widetilde_q} \\[2mm]
            & \bullet \quad \widetilde{w} := 2 \, \check{Q}_{V, 12} \, \widetilde{c}_{2,3} + \widetilde{w}_{V,1} \; ; \label{eq:widetilde_w} \\[2mm]
            & \bullet \quad \widetilde{r} := \widetilde{c}_{2,3}^{\ T} \, \check{Q}_{V, 22} \, \widetilde{c}_{2,3} + \widetilde{w}_{V,2}^{\ T} \, \widetilde{c}_{2,3} + r T \; . \label{eq:widetilde_r}
        \end{align}
    \end{itemize}
\end{proof}

The optimisation problem \eqref{eq:opt_ref_tilde} is then equivalent to the following one in the present quadratic setting:
\begin{equation} \label{eq:opt_ref_quad}
    \left\{
    \begin{array}{l}
         \displaystyle \widetilde{c}_1^\star \in \argmin_{\widetilde{c}_1 \in \R^\sigma} \widetilde{c}_1^T \, \widetilde{Q} \, \widetilde{c}_1 + \widetilde{w}^T \widetilde{c}_1 + \kappa \sum_{i=1}^I \omega_i \, \big( \widetilde{c}_1 - \widetilde{c}_{R_i, 1} \big)^T \Lambda_{\Sigma,1}^{-1} \big( \widetilde{c}_1 - \widetilde{c}_{R_i, 1} \big)  \\[2mm]
         \widetilde{c}_2 = V_2^T c_{R_i} \\[2mm]
         \widetilde{c}_3 = S_{A,2}^{-1} \, U^T \Gamma
    \end{array} \; .
    \right.
\end{equation}
In the following result, we provide a sufficient condition on the parameter $\kappa > 0$ so that the problem \eqref{eq:opt_ref_quad} is a quadratic program. The proof uses the fact that the symmetric matrix associated with the quadratic objective function is now explicit and given by the sum of two matrices. A perturbation result for matrices is then applied to obtain a bound for $\kappa$ assuring that the symmetric matrix is positive semidefinite.

\begin{theorem} \label{thm:quad_prog}
    Let $\rho_1 \geqslant \rho_2 \geqslant \dots \geqslant \rho_\sigma$ and $\lambda_1 \geqslant \lambda_2 \geqslant \dots \geqslant \lambda_K$ be respectively the eigenvalues of the symmetric matrices $\widetilde{Q}$ and $\Sigma$. If $\kappa \geqslant - \rho_\sigma \, \lambda_1$ then the optimisation problem \eqref{eq:opt_ref_quad} is a quadratic program.
\end{theorem}
\begin{proof}
    We first note that all the eigenvalues of the matrix $\Sigma$ are non-negative (because $\Sigma$ is a covariance matrix) and that $\lambda_{\sigma + 1} = \ldots = \lambda_K = 0$ (because $\rk \Sigma = \sigma$). In particular, the eigenvalue $\lambda_1$ is positive.\\
    Standard calculations show that the symmetric matrix associated with the quadratic objective function of the problem \eqref{eq:opt_ref_quad} is given by
    \begin{equation*}
        M(\kappa) := \widetilde{Q} + \kappa \, \Lambda_{\Sigma, 1}^{-1} \in \R^{\sigma \times \sigma} \; .
    \end{equation*}
    Let $\mu_1(\kappa) \geqslant \mu_2(\kappa) \geqslant \dots \geqslant \mu_\sigma(\kappa)$ denote the eigenvalues of $M(\kappa)$. The goal is then to find a sufficient condition on $\kappa > 0$ so that $\mu_\sigma(\kappa)$ is non-negative to assure that $M$ is positive semidefinite. Since $M(\kappa)$ can be interpreted as a perturbed version of $\widetilde{Q}$, we can apply Weyl's inequality (see for instance \citet{wang201965}) which states
    \begin{equation*}
        \mu_\sigma(\kappa) \geqslant \rho_\sigma + \frac{\kappa}{\lambda_1} \; .
    \end{equation*}
    Then choosing $\kappa$ such that $\kappa \geqslant - \rho_\sigma \, \lambda_1$ implies that $\mu_\sigma(\kappa) \geqslant 0$, leading to the result.
\end{proof}

For the sake of completeness, we finish by rewriting the problem \eqref{eq:opt_ref_quad} as a quadratic optimisation problem in $\mathcal{V}_1 \subset \R^K$.

\begin{proposition}
    Suppose that the cost function $F$ is of the form \eqref{eq:quad_F}. Then the optimisation problem \eqref{eq:opt_ref} is equivalent to the following one:
    \begin{equation*}
        c^\star \in \argmin_{c \in \mathcal{V}_1} c^T \Big( \check{Q} + \kappa \, \Sigma^\dagger \Big) c + \left( \check{w} - 2 \kappa \sum_{i=1}^I \omega_i \, \Sigma^\dagger c_{R_i} \right)^{\hspace{-5pt}T} c \; .
    \end{equation*}
\end{proposition}
\begin{proof}
    It is sufficient to show that the two following objective functions $g_1, g_2: \R^K \longrightarrow \R$ have the same minima:
    \begin{itemize}
        \item $\displaystyle g_1(c) := \check{F}(c) + \kappa \sum_{i=1}^I \omega_i \, \big( c - c_{R_i} \big)^T \Sigma^\dagger \big( c - c_{R_i} \big)$ ;
        \item $\displaystyle g_2(c) := c^T \Big( \check{Q} + \kappa \, \Sigma^\dagger \Big) c + \left( \check{w} - 2 \kappa \sum_{i=1}^I \omega_i \, \Sigma^\dagger c_{R_i} \right)^{\hspace{-5pt}T} c$ .
    \end{itemize}
    Firstly we have by standard calculations,
    \begin{align*}
        \sum_{i=1}^I \omega_i \, \big( c - c_{R_i} \big)^T \Sigma^\dagger \big( c - c_{R_i} \big)
        & = \sum_{i=1}^I \omega_i \, c^T \Sigma^\dagger c - 2 \sum_{i=1}^I \omega_i \, c_{R_i}^{\ T} \Sigma^\dagger c + \sum_{i=1}^I \omega_i \, c_{R_i}^{\ T} \Sigma^\dagger c_{R_i} \\
        & = c^T \Sigma^\dagger c - \left( 2 \sum_{i=1}^I \omega_i \, \Sigma^\dagger  c_{R_i} \right)^{\hspace{-5pt}T} c + \sum_{i=1}^I \omega_i \, c_{R_i}^{\ T} \Sigma^\dagger c_{R_i} \; ,
    \end{align*}
    for any $c \in \R^K$, where we have used $\sum_{i=1}^I \omega_i = 1$. Combining now this equality with \cref{lem:ff} implies
    \begin{align*}
        g_1(c)
            & = c^T \check{Q} c + \check{w}^T c + r T + \kappa \left( c^T \Sigma^\dagger c - \left( 2 \sum_{i=1}^I \omega_i \, \Sigma^\dagger  c_{R_i} \right)^{\hspace{-5pt}T} c + \sum_{i=1}^I \omega_i \, c_{R_i}^{\ T} \Sigma^\dagger c_{R_i} \right) \\
            & = c^T \Big( \check{Q} + \kappa \, \Sigma^\dagger \Big) c + \left( \check{w} - 2 \kappa \sum_{i=1}^I \omega_i \, \Sigma^\dagger c_{R_i} \right)^{\hspace{-5pt}T} c + \kappa \sum_{i=1}^I \omega_i \, c_{R_i}^{\ T} \Sigma^\dagger c_{R_i} + r T \\
            & = g_2(c) + \kappa \sum_{i=1}^I \omega_i \, c_{R_i}^{\ T} \Sigma^\dagger c_{R_i} + r T \; .
    \end{align*}
    Since the two last terms of the last right-hand side do not depend on $c$, we deduce that the objective functions $g_1$ and $g_2$ have the same minima.
\end{proof}

\bibliography{biblio}

\end{document}